\documentstyle[12pt,epsf,epsfig]{article}
\newcount\timecount
\newcount\hours \newcount\minutes  \newcount\temp \newcount\pmhours

\hours = \time
\divide\hours by 60
\temp = \hours
\multiply\temp by 60
\minutes = \time
\advance\minutes by -\temp
\def\hour{\the\hours}
\def\minute{\ifnum\minutes<10 0\the\minutes
            \else\the\minutes\fi}
\def\clock{
\ifnum\hours=0 12:\minute\ AM
\else\ifnum\hours<12 \hour:\minute\ AM
      \else\ifnum\hours=12 12:\minute\ PM
            \else\ifnum\hours>12
                 \pmhours=\hours
                 \advance\pmhours by -12
                 \the\pmhours:\minute\ PM
                 \fi
            \fi
      \fi
\fi
}

\def\monthname{\relax\ifcase\month 0/\or January\or February\or
   March\or April\or May\or June\or July\or August\or September\or
   October\or November\or December\else\number\month/\fi}

\def\bold#1{\setbox0=\hbox{$#1$}%
     \kern-.025em\copy0\kern-\wd0
     \kern.05em\copy0\kern-\wd0
     \kern-.025em\raise.0433em\box0 }

\def\ga{\mathrel{\raise.3ex\hbox{$>$\kern-.75em\lower1ex\hbox{$\sim$}}}}
\def\la{\mathrel{\raise.3ex\hbox{$<$\kern-.75em\lower1ex\hbox{$\sim$}}}}
\def\gev{{\rm \, Ge\kern-0.125em V}}
\def\tev{{\rm \, Te\kern-0.125em V}}
\def\beq{\begin{equation}}
\def\eeq{\end{equation}}

\def\m12{m_{1\!/2}}

\def\Yi{\eta^{\ast}_{11} \left( \frac{y_{i}}{2} g' Z_{\chi 1} + 
        g T_{3i} Z_{\chi 2} \right) + \eta^{\ast}_{12} 
        \frac{g m_{q_{i}} Z_{\chi 5-i}}{2 m_{W} B_{i}}}

\def\Xii{\eta^{\ast}_{11} 
        \frac{g m_{q_{i}}Z_{\chi 5-i}^{\ast}}{2 m_{W} B_{i}} - 
        \eta_{12}^{\ast} e_{i} g' Z_{\chi 1}^{\ast}}

\def\Wi{\eta_{21}^{\ast}
        \frac{g m_{q_{i}}Z_{\chi 5-i}^{\ast}}{2 m_{W} B_{i}} -
        \eta_{22}^{\ast} e_{i} g' Z_{\chi 1}^{\ast}}

\def\Vi{\eta_{22}^{\ast} \frac{g m_{q_{i}} Z_{\chi 5-i}}{2 m_{W} B_{i}}
        + \eta_{21}^{\ast}\left( \frac{y_{i}}{2} g' Z_{\chi 1}
        + g T_{3i} Z_{\chi 2} \right)}

\def\zthree{\delta_{1i} [g Z_{\chi 2} - g' Z_{\chi 1}]}

\def\zfour{\delta_{2i} [g Z_{\chi 2} - g' Z_{\chi 1}]}

\begin{document}
\begin{titlepage}
\pagestyle{empty}
\baselineskip=21pt
\rightline{hep-ph/0007113}
\rightline{CERN--TH/2000-188}
\rightline{UMN--TH--1914/00}
\rightline{TPI--MINN--00/36}
\vskip 0.2in
\begin{center}
{\large{\bf Exploration of Elastic Scattering Rates for Supersymmetric
Dark Matter}}
\end{center}
\begin{center}
\vskip 0.2in
{{\bf John Ellis}$^1$, {\bf Andrew Ferstl}$^2$ and {\bf Keith
A.~Olive}$^{2,3}$}\\
\vskip 0.1in
{\it
$^1${TH Division, CERN, Geneva, Switzerland}\\
$^2${School of Physics and Astronomy,
University of Minnesota, Minneapolis, MN 55455, USA}\\
$^3${Theoretical Physics Institute,
University of Minnesota, Minneapolis, MN 55455, USA}}\\
\vskip 0.2in
{\bf Abstract}
\end{center}
\baselineskip=18pt \noindent
We explore the possible cross sections for the elastic scattering of
neutralinos $\chi$ on nucleons $p,n$ in the minimal supersymmetric 
extension of the standard model (MSSM). 
Universality of the soft supersymmetry-breaking scalar masses for the
Higgs multiplets is not assumed, but the MSSM parameters are
nevertheless required to lead consistently to an electroweak vacuum.
We explore
systematically the region of MSSM parameter space where LEP and
other accelerator constraints are respected, and the relic neutralino
density lies in the range $0.1 \le \Omega_{\chi} h^2 \le 0.3$ preferred by
cosmology. We also discuss models with
$\Omega_{\chi} h^2 < 0.1$, in which case we scale the density of
supersymmetric dark matter in our galactic halo by $\Omega_{\chi} h^2 /
0.1$, allowing for the possible existence of some complementary form
of cold dark matter.
We find values of the cross sections that are
considerably lower than the present experimental sensitivities.
At low neutralino masses, $m_\chi \la 100$ GeV, the cross sections 
may be somewhat higher than in the constrained MSSM with universal soft
Higgs masses, though they are generally lower.
In the case of large $m_\chi$, the cross sections we find may
be considerably larger than in the constrained model, but still
well below the present experimental sensitivity.
\vfill
\leftline{CERN--TH/2000-188}
\leftline{July 2000}
\end{titlepage}
\baselineskip=18pt
\section{Introduction}

One of the key issues at the frontier between particle physics and
cosmology is the nature of the non-baryonic dark matter that
apparently dominates the matter density of the Universe.
This is probably dominated by cold dark matter, with a density
that probably falls within the range $0.2 <
\Omega_{CDM} < 0.5$~\cite{triangle}, and may be in
the form of massive weakly-interacting particles. It is therefore
particularly important to search for such dark matter
particles~\cite{GJK},
and one of the most direct strategies is the search for relic
particle scattering on nuclei in a laboratory detector~\cite{GW}. Many
experiments around the world are engaged in this search, largely
motivated by the cross sections calculated assuming that the cold
dark matter is dominated by the lightest neutralino $\chi$~\cite{EHNOS} of
the minimal
supersymmetric extension of the Standard Model (MSSM)~\cite{MSSM}.

We recently re-evaluated~\cite{EFlO1} the spin-dependent and
spin-independent
cross sections for neutralino scattering on protons and
neutrons~\cite{EF,etal},
assuming universality for all of the soft supersymmetry-breaking mass
parameters of the MSSM including the Higgs multiplets, incorporating the
latest available LEP constraints
on the MSSM parameter space, and assuming that the
cosmological density of the relic neutralino falls within the range $0.1 <
\Omega_\chi h^2 < 0.3$, corresponding to the favoured range of
$\Omega_{CDM}$ and a Hubble expansion rate $0.6 \le h \le 0.8$
in units $H_0 \equiv 100 \times h$~km/s/Mpc. We used the latest
information from chiral symmetry~\cite{leut,Cheng}, low-energy $\pi -
p,n$
scattering~\cite{Gasser} and deep-inelastic lepton-nucleon
scattering~\cite{Mallot} to fix the hadronic matrix elements.
Our calculations
fell considerably below the present experimental
sensitivities~\cite{Gait},
as well as the highest theoretical estimates available in the
literature~\cite{bigguys}, some of which used less restrictive
assumptions. There are, however, some other recent lower estimates: 
see~\cite{lowguys}, for example, which is in good agreement with our
previous work~\cite{EFlO1}.

Shortly after our paper appeared, the DAMA collaboration
confirmed~\cite{DAMA} their previous evidence for the annual modulation of
energy deposits in their scintillation detector, which they
interpret as due to the scattering of some cold dark matter
particle with mass between about 50 and 100~GeV, and
spin-independent cross section on a proton between about $10^{-6}$ and
$10^{-5}$~pb. This cross section range is considerably larger than
we found previously~\cite{EFlO1}, though consistent with the range allowed
by some
previous cross section estimates. Subsequent to the DAMA paper,  
the CDMS collaboration has reported~\cite{CDMS} negative results from
their
experiment, establishing an upper limit on the spin-independent 
cross section that excludes most, but not all, of the range
suggested by DAMA.

This unresolved situation motivates us to explore more widely
the possible neutralino-proton cross sections in the MSSM,
including both the spin-dependent and spin-independent (scalar)
contributions. As before, we impose the latest constraints on
the MSSM parameter space imposed by the LEP and other
experiments~\cite{EFGOS,EFGO},
such as measurements of $b \rightarrow s \gamma$ decay. 
It is important to note that the LEP limits we use here~\cite{EFGO} have
been
updated significantly compared to what we used in~\cite{EFlO1}. Notably,
the chargino and particularly the Higgs mass limits we use here
are stronger. The latter has a substantial effect at $\tan\beta =
3$: in addition to the consequent direct reduction in
the Higgs-exchange contribution to the scalar cross section, the improved
lower limit on the Higgs mass further restricts $m_0$ and $m_{1/2}$
from below, because
of their contribution to $m_h$ via radiative corrections.
Also, previously we did not use the $b \rightarrow s \gamma$ constraint,
which we implement here by
requiring $m_A > 300$~GeV for $\mu < 0$. All of these effects tend to
remove some of the higher cross
sections that we found previously, particularly at low $m_\chi$.

The main thrust of this paper, however, is to
relax two of the theoretical assumptions made in our
previous work. 

$\bullet$
The absence of large flavour-changing neutral interactions
suggests that the soft super- symmetry-breaking scalar mass parameters
$m_{0_i}$
of the MSSM may be universal for different quark and lepton flavours.
However, there is no strong phenomenological or theoretical
reason why the $m_{0_i}$ should be the same for the Higgs multiplets as
for squarks and sleptons, and we relax this universality assumption
in this work. It is known that, in this case, the lightest
neutralino $\chi$ might be mainly a Higgsino, but this particular
option is greatly restricted by LEP data~\cite{EFGOS,EFGO}.

$\bullet$
Neutralinos might not constitute all the cold dark matter, but might
be complemented by other particles such as axions or superheavy relics.
In this case, $\Omega_\chi < \Omega_{CDM}$, and $\Omega_\chi h^2 < 0.1$
becomes a possibility. For any given neutralino mass, $\Omega_\chi$
may be decreased by increasing the $\chi$ annihilation cross
sections, which is often correlated with an enhanced elastic $\chi$-proton
scattering cross section. Before concluding that cold dark matter
detection becomes easier in this case, however, one must consider
what fraction of our galactic halo density $\rho_{halo}$
could be composed of neutralinos.
Since the process of halo formation is essentially independent of the
nature of the cold dark matter, as long as it is non-relativistic and
weakly interacting, one should expect that
\begin{equation}
\rho_\chi = \rho_{halo} \times 
\left( {\Omega_\chi \over \Omega_{CDM}} \right),
\label{halofraction}
\end{equation}
In an effort to be as optimistic as is reasonable, we assume that
$\rho_\chi = \rho_{halo}$ if $\Omega_\chi h^2 \ge 0.1$, and rescale:
$\rho_\chi = \rho_{halo} \times (\Omega_\chi h^2 / 0.1)$ if
$\Omega_\chi h^2 \le 0.1$.

In our previous work~\cite{EFlO1}, in which we assumed universality for
the Higgs
masses (UHM) 
at the conventional supersymmetric GUT scale $\sim 10^{16}$~GeV,
and the canonical range $0.1 < \Omega_\chi h^2 < 0.3$,
we found that the possible ranges of elastic scattering cross
sections were very narrow for any fixed values of $m_\chi, \tan\beta$
and the sign of $\mu$, even allowing for plausible uncertainties in
the hadronic inputs~\cite{leut,Cheng,Gasser,Mallot},
and that they were always orders of magnitude
below the present sensitivities~\cite{Gait}, even for the smallest allowed
values
of $m_\chi \sim 50$~GeV~\cite{EFGOS,EFGO}.
Specifically, the maximum value we found for the spin-dependent $\chi - p$
elastic scattering cross section for $3 \le \tan\beta \le 10$ was
well below $10^{-3}$~pb, attained for $m_\chi \sim 60$~GeV, and
the maximum value we found for the spin-independent $\chi - p$
elastic scattering cross section for $3 \le \tan\beta \le 10$ was
$\sim 10^{-7}$~pb, again attained for $m_\chi \sim 60$~GeV.
The corresponding experimental sensitivities are $\sim 1$~pb and $\sim 3
\times 10^{-6}$~pb, respectively. At higher neutralino masses, the
predicted cross-sections were significantly smaller still. 

In the constrained version of the MSSM, when all soft scalar masses,
including the Higgs masses, are set equal at the unification scale (UHM),
there are four independent parameters, the soft scalar masses, $m_0$, the
gaugino masses, $m_{1/2}$, the soft trilinear mass terms, $A$ (assumed to
be universal), and $\tan \beta$. In addition, there is the freedom to
choose the sign of the Higgs mixing mass $\mu$.  Previously we scanned
the $m_0 - m_{1/2}$ parameter space for fixed $\tan \beta$ and
$sgn(\mu)$. Our
results were not very sensitive to $A$.  

Now that we relax the universal
Higgs-mass assumption (nUHM), we find much broader ranges of elastic
scattering cross sections for any fixed values of
$m_\chi, \tan\beta$ and the sign of $\mu$. As previously,
we perform a systematic scan of the region of the $m_0, m_{1/2}$
parameter space of the MSSM that is
consistent with accelerator constraints. 
Here, $m_0$ refers only
to a common squark and slepton mass, and the two Higgs soft masses
$m_1$ and $m_2$ are fixed by the conditions of electroweak
symmetry breaking, since we allow $\mu$ and the Higgs pseudoscalar
mass $m_A$ to be free parameters. Thus, we scan over $m_0, m_{1/2}, \mu,
m_A$,
and $A$ for fixed $\tan \beta$. The details of these scans are given
below, where we document which parameter choices fail which LEP
constraint and/or the cosmological relic density requirement.  

We find that the elastic scattering cross sections may
be somewhat larger than we found before in the UHM case, particularly for
larger
$m_\chi$. However, the absolute values are still well below the present
experimental sensitivities~\cite{Gait}, at least
for the canonical range $0.1 < \Omega_\chi h^2
< 0.3$ for the relic neutralino density. This remains true when
we consider $ \Omega_\chi h^2 < 0.1$, but rescale the halo density
as described above.

We cannot exclude the possibility that there might be some
variant of the MSSM that could accommodate the cold dark matter
scattering interpretation of the DAMA data, but this would require
an extension of the framework discussed here. One possibiliity
might be to adopt a larger value of $\tan\beta$~\cite{arno}: we restrict
our attention to $\tan\beta \le 10$ 
to avoid some uncertainties in the treatment of radiative corrections in
the renormalization-group evolution
of the MSSM parameters which affect the relic density
calculations. 
Another possibility might be to relax
further the universality assumptions for soft supersymmetry-breaking
masses, either in the scalar or the gaugino sector. In particular, models
in which $m_{\tilde q} / m_{\tilde
\ell}$ is smaller than in the models discussed here might be able to
accommodate larger elastic $\chi$-proton rates for any given value of
$\Omega_{\chi}$. Another way to reduce $m_{\tilde q} / m_{\tilde \ell}$,
with a similar effect, could be to postulate universality at a lower,
intermediate renormalization scale, below the conventional supersymmetric
GUT scale
\cite{lowuni}.

\section{Theoretical and Phenomenological Background}

We review in this Section relevant aspects
of the MSSM~\cite{MSSM}. The neutralino LSP is the
lowest-mass eigenstate combination of the Bino ${\tilde B}$, Wino $\tilde W$
and Higgsinos ${\tilde H}_{1,2}$, whose mass matrix $N$ is
diagonalized by a matrix $Z$: $diag(m_{\chi_1,..,4}) = Z^* N Z^{-1}$.
The composition of the lightest neutralino may be written as
\begin{equation}
\chi = Z_{\chi 1}\tilde{B} + Z_{\chi 2}\tilde{W} +
Z_{\chi 3}\tilde{H_{1}} + Z_{\chi 4}\tilde{H_{2}}
\label{id}
\end{equation}
As previously, we neglect CP violation in this paper, so that
there are no CP-violating phases in the neutralino
mass matrix and mixing. For the effects of CP-violating phases on the
neutralino scattering cross-section see \cite{FFO1}-\cite{fg}. We assume
universality at the supersymmetric GUT scale for the 
$U(1)$ and $SU(2)$ gaugino masses: $M_{1,2} = m_{1/2}$, so that 
$M_1 = \frac{5}{3}\tan^2{\theta_{W}}M_{2}$ at the electroweak scale.

We also assume
GUT-scale universality for the soft supersymmetry-breaking scalar masses
$m_0$ of the squarks and sleptons, but {\bf NOT}
for the Higgs bosons, in contrast to~\cite{EFlO1}.
We further
assume GUT-scale universality for the soft supersymmetry-breaking
trilinear terms $A$.
Our treatment of the sfermion mass matrices $M$ follows
those in~\cite{FFO1}, and we refer the interested reader to~\cite{EFlO1}
for further details and notation. It suffices here to recall that,
CP being conserved,
the sfermion mass-squared matrix for each flavour $f$ 
is diagonalized by a
rotation through an angle $ \theta_{f} $.
We treat as free parameters $m_{1/2}$ (we actually use $M_2$ which is
equal to $m_{1/2}$ at the unification scale), the soft
supersymmetry-breaking scalar mass scale
$m_{0}$ (which in the present context refers only to the universal
sfermion masses at the unification scale),
$A$ and $\tan\beta$. In addition, we treat $\mu$ and the pseudoscalar
Higgs mass
$m_A$ as independent parameters, and thus the two Higgs soft
masses $m_1$ and $m_2$, are specified by the electroweak vacuum
conditions, which we calculate using $m_t = 175$~GeV~\footnote{We
have checked that varying $m_t$ by $\pm 5$~GeV has a negligible
effect on our results.}. 

The MSSM Lagrangian leads to the following low-energy effective
four-fermion
Lagrangian suitable for describing elastic $\chi$-nucleon
scattering~\cite{FFO1}:
\begin{equation}
{\cal L} = \bar{\chi} \gamma^\mu \gamma^5 \chi \bar{q_{i}} 
\gamma_{\mu} (\alpha_{1i} + \alpha_{2i} \gamma^{5}) q_{i} +
\alpha_{3i} \bar{\chi} \chi \bar{q_{i}} q_{i} + 
\alpha_{4i} \bar{\chi} \gamma^{5} \chi \bar{q_{i}} \gamma^{5} q_{i}+
\alpha_{5i} \bar{\chi} \chi \bar{q_{i}} \gamma^{5} q_{i} +
\alpha_{6i} \bar{\chi} \gamma^{5} \chi \bar{q_{i}} q_{i}
\label{lagr}
\end{equation}
This Lagrangian is to be summed over the quark generations, and the 
subscript $i$ labels up-type quarks ($i=1$) and down-type quarks
($i=2$).  The terms with coefficients $\alpha_{1i}, \alpha_{4i},
\alpha_{5i}$ and $\alpha_{6i}$ make contributions to the 
elastic scattering cross section that are velocity-dependent,
and may be neglected for our purposes. In fact, if the 
CP-violating phases are absent as assumed here, $\alpha_5 = \alpha_6 =
0$~\cite{FFO2,CIN}. The coefficients relevant for our discussion are: 

\begin{eqnarray}
\alpha_{2i} & = & \frac{1}{4(m^{2}_{1i} - m^{2}_{\chi})} \left[
\left| Y_{i} \right|^{2} + \left| X_{i} \right|^{2} \right] 
+ \frac{1}{4(m^{2}_{2i} - m^{2}_{\chi})} \left[ 
\left| V_{i} \right|^{2} + \left| W_{i} \right|^{2} \right] \nonumber \\
& & \mbox{} - \frac{g^{2}}{4 m_{Z}^{2} \cos^{2}{\theta_{W}}} \left[
\left| Z_{\chi_{3}} \right|^{2} - \left| Z_{\chi_{4}} \right|^{2}
\right] \frac{T_{3i}}{2}
\label{alpha2}
\end{eqnarray}
and
\begin{eqnarray}
\alpha_{3i} & = & - \frac{1}{2(m^{2}_{1i} - m^{2}_{\chi})} Re \left[
\left( X_{i} \right) \left( Y_{i} \right)^{\ast} \right] 
- \frac{1}{2(m^{2}_{2i} - m^{2}_{\chi})} Re \left[ 
\left( W_{i} \right) \left( V_{i} \right)^{\ast} \right] \nonumber \\
& & \mbox{} - \frac{g m_{qi}}{4 m_{W} B_{i}} \left[ Re \left( 
\zthree \right) D_{i} C_{i} \left( - \frac{1}{m^{2}_{H_{1}}} + 
\frac{1}{m^{2}_{H_{2}}} \right) \right. \nonumber \\
& & \mbox{} +  Re \left. \left( \zfour \right) \left( 
\frac{D_{i}^{2}}{m^{2}_{H_{2}}}+ \frac{C_{i}^{2}}{m^{2}_{H_{1}}} 
\right) \right]
\label{alpha3}
\end{eqnarray}
where
\begin{eqnarray}
X_{i}& \equiv& \Xii \nonumber \\
Y_{i}& \equiv& \Yi \nonumber \\
W_{i}& \equiv& \Wi \nonumber \\
V_{i}& \equiv& \Vi
\label{xywz}
\end{eqnarray}
where $y_i, T_{3i}$ denote hypercharge and isospin, and
\begin{eqnarray}
\delta_{1i} = Z_{\chi 3} (Z_{\chi 4}) &,& \delta_{2i} = Z_{\chi 4}
(-Z_{\chi 3}), \nonumber \\
B_{i} = \sin{\beta} (\cos{\beta}) &,& A_{i} = \cos{\beta} ( -\sin{\beta}), 
\nonumber \\
C_{i} = \sin{\alpha} (\cos{\alpha}) &,& D_{i} = \cos{\alpha} ( -
\sin{\alpha}) 
\label{moredefs}
\end{eqnarray}
for up (down) type quarks. We denote by $m_{H_2} < m_{H_1}$
the two scalar Higgs masses, and $ \alpha $ denotes the Higgs mixing
angle~\footnote{We note that (\ref{alpha2}, \ref{alpha3}) is taken from
\cite{FFO2}
and agree with~\cite{GJK,EF,CIN}.}.

As discussed in~\cite{EFlO1},
the elastic cross section for scattering off a nucleus can be
decomposed into a scalar (spin-independent) part obtained from the
$\alpha_{2i}$ term in (\ref{lagr}), and a spin-dependent part
obtained from the $\alpha_{3i}$ term. Each of these can be
written in terms of the cross sections for elastic scattering
for scattering off individual nucleons.
We re-evaluated the relevant matrix elements in~\cite{EFlO1}.
Here we limit ourselves to recalling that:

$\bullet$
There are uncertainties in the {\it scalar part} of the
cross section associated with
the ratios of the light-quark masses, which we take from~\cite{leut}:
\beq
{m_u \over m_d} = 0.553 \pm 0.043 , \qquad
{m_s \over m_d} = 18.9 \pm 0.8
\eeq
and information from chiral symmetry applied to baryons.
Here the principal uncertainty is associated with the experimental value
of the $\pi$-nucleon $\sigma$ term
and the corresponding values of the ratios of the $B_q \equiv <p|{\bar q}q
| p >$. Following~\cite{Cheng}, we use
\beq
z \equiv {B_u - B_s \over B_d - B_s} = 1.49
\label{chengvalue}
\eeq
with a negligible experimental error, and~\cite{Gasser}
\beq
y \equiv {2 B_s \over B_d + B_u} = 0.2 \pm 0.1,
\label{defy}
\eeq
which yields
\beq
{B_d \over B_u} = 0.73 \pm 0.02
\eeq
The difference between the scalar parts
of the cross sections for scattering off protons and neutrons are
rather small.

$\bullet$
The spin-dependent part of the elastic $\chi$-nucleus cross section can be
written in terms of axial-current matrix elements
$\Delta_{i}^{(p,n)}$ that parametrize the quark spin content of the
nucleon. We extract from a recent global analysis~\cite{Mallot}
the values
\beq
\Delta_{u}^{(p)} = 0.78 \pm 0.04, \qquad \Delta_{d}^{(p)} = -0.48 \pm
0.04,
\qquad \Delta_{s}^{(p)} = - 0.15 \pm 0.04
\label{spincontent}
\eeq
where the errors are essentially 100\%
correlated
for the three quark flavours. In the case of the neutron, we have
$\Delta_{u}^{(n)} = \Delta_{d}^{(p)}, 
\Delta_{d}^{(n)} = \Delta_{u}^{(p)}$, and 
$\Delta_{s}^{(n)} = \Delta_{s}^{(p)}$.

\section{Cosmological and Experimental Constraints}

As noted in~\cite{EFlO1}, several convergent measures of
cosmological parameters~\cite{triangle} have suggested that the cold dark
matter
density $\Omega_{CDM} = 0.3 \pm 0.1$ and that the Hubble
expansion rate $H \equiv h \times 100$~km/s/Mpc: $h = 0.7 \pm 0.1$,   
leading to our preferred range $0.1 \le \Omega_{CDM} h^2 \le 0.3$.
The recent data on the spectrum of cosmic microwave background
fluctuations from BOOMERANG~\cite{BOOM} and MAXIMA~\cite{MAXI} are
consistent with this
range, but do not significantly constrain it further.
The upper limit on $\Omega_{CDM}$ can be translated directly into
the corresponding upper limit on $\Omega_\chi$. However, it is 
possible that there is more than one component in the cold dark
matter, for example axions and/or superheavy relics as well as
the LSP $\chi$, opening up the
possibility that $\Omega_\chi < 0.1$.
For a given value of $m_\chi$, values of the MSSM parameters which lead to
$\Omega_\chi < 0.1$ tend to have larger $\chi$ annihilation
cross sections, and hence
larger elastic scattering cross sections. Note, however, that the upper
bound, $\Omega_\chi h^2 < 0.3$, is a firm upper bound relying only on
the lower limit to the age of the Universe, $t_U > 1.2 \times 10^{10}$
years (with $\Omega_{\rm total} \le 1$). 

However, in such a `shared'
cold dark matter scenario, the packing fraction of neutralinos in the
galactic halo must be reduced. As discussed in~\cite{ES}, for
example, dark matter particles are taken into the halo in `sheets' in
phase space, whose thicknesses are determined by their initial (thermal)
velocity. The `sheets' of cold dark matter particles are of negligible
thickness, so the ratios of their densities in the halo are identical with
their cosmological densities, and therefore
\begin{equation}
{ \rho_\chi \over \rho_{CDM} } = { \Omega_\chi \over \Omega_{CDM} }
\label{CDMratio}
\end{equation}
On the other hand, the `sheets'
of hot dark matter particles are of finite thickness related to their
thermal velocities at the onset of structure formation,
which limits the possible phase-space density of hot dark matter
particles, so that $\rho_{HDM} / \rho_{CDM} < \Omega_{HDM} /
\Omega_{CDM}$ in general~\cite{ES}. Moreover, a large ratio $\Omega_{HDM}
/ \Omega_{CDM}$ is currently not expected.

The LSP detection
rate also must be reduced correspondingly to (\ref{CDMratio}).
Accordingly, when we consider MSSM parameter choices that
have $\Omega_\chi h^2 \le 0.1$, we rescale the calculated
scattering rate by a factor $\Omega_\chi h^2 / 0.1$. This rescaling
by the minimal acceptable value of $\Omega_{CDM} h^2$ is relatively
optimistic.

For the calculation of the relic LSP density, we have included radiative
corrections \cite{EFGOS} to the neutralino mass matrix and include all
possible annihilation channels \cite{McOS}.  In the MSSM, it is well
known that there are large regions of the $M_2, \mu$ parameter plane for
which the LSP and the next lightest neutralino (NLSP) and/or chargino are
nearly degenerate, namely in the Higgsino portion of the plane when 
$M_2 \gg \mu$. It was shown~\cite{gs,co2} that, in these regions,
coannihilations between the LSP, NLSP, and charginos are of particular
importance in determining the final relic density of LSPs, and 
these have been included in the present
calculation. Inclusion of these coannihilation
channels has the important consequence that, in the Higgsino
regions where one expects larger elastic scattering cross sections, the
relic abundance is substantially reduced. On the other hand, we do not
include here
coannihilations between the LSP and the sleptons $\tilde \ell$
\cite{EFOSi}, in particular the lighter stau $\tilde \tau_1$, 
which were shown
to play an important role in models with scalar mass universality also
for the Higgs multiplets (UHM). These are known, in particular, to be
important for determining the maximum possible
generic value of $m_\chi$ in the UHM case, but are of less generic 
importance than $\chi - \chi' -  \chi^\pm$ coannihilations in the
non-universal nUHM case considered here. For the same reason, we have also
not
implemented $\chi - \tilde t$ coannihilation~\cite{BDD}. 
The neglect of such
$\chi - \tilde f$ coannihilation processes is generally conservative 
as far as the elastic
scattering rates are concerned, since any reduction they cause in
$\Omega_\chi h^2$ is unlikely to be compensated by a corresponding 
enhancement in the elastic
scattering cross section. We also do not pay any particular attention to
the narrow parameter slice of mixed gaugino/Higgsino dark matter where
$|\mu| \propto m_{1/2}$ and
$m_\chi$ may become large~\cite{FMW}, because this requires an adjustment
of parameters at the \% level, and is hence not generic. However, these
are sampled, with the appropriate weighting, in our general randomized
scan of the parameter space.

The lower limit on $m_\chi$ depends on the
sparticle search limits provided by LEP and other
experiments~\cite{EFGOS,EFGO}. The most essential
of these
for our current purposes are those provided by the experimental
lower limits on the lighter chargino mass $m_{\chi^\pm}$ and the
lighter scalar Higgs mass $m_{H_2}$. As discussed in~\cite{EFGO},
here we assume a lower limit $m_{\chi^\pm} \ge 101$~GeV.
The impact of the recently-improved lower
limit on the Higgs mass~\cite{newLEP} is potentially more
significant~\cite{EFGO},
particularly
for $\tan\beta = 3$, as displayed in Figs.~6 of~\cite{EFGO}. The
present experimental lower limit for $\tan\beta = 3$ approaches
$m_{H_2} > 107$~GeV~\cite{newLEP}. In implementing this constraint,
we allow a safety margin of $\sim 3$~GeV in the MSSM calculations of
$m_{H_2}$~\cite{newHiggs}, and hence require the MSSM calculation to
yield $m_{H_2} > 104$~GeV for $\tan\beta = 3$. In the case of
$\tan\beta = 10$, the LEP constraint on the MSSM Higgs mass is weaker
(see Fig.~6 of~\cite{EFGO}), and we require only $m_{H_2} > 86 $~GeV,
which includes again a 3~GeV margin of uncertainty. The corresponding
limit on
$m_0$ and $m_{1/2}$ in this case may be ignored~\cite{EFGO}.
The other two constraints that we implement are on sfermion masses, 
which we require to be (i) larger than 92 GeV, and (ii) larger than
that of the lightest neutralino. We recall also the importance of the $b
\rightarrow s
\gamma$ constraint~\cite{bsg}, which we
implement in an approximate way, by requiring $m_A >
300$~GeV for $\mu < 0$~\cite{EFGO}.  
As also discussed in~\cite{EFGO}, requiring our
present
electroweak vacuum to be stable against transitions to a lower-energy
state in which electromagnetic charge and colour are broken
(CCB)~\cite{AF} would remove a large part of the cosmologically-favoured
domain of MSSM parameter space. We have not implemented this 
optional requirement
in the present study.  In the next section, we will show the effect of the
various expermental constraints on our scan of the parameter plane.

\section{MSSM Parameter Scan}

We have scanned systematically the MSSM parameter space, taking
into account the cosmological and experimental constraints
enumerated in the previous Section and implementing the MSSM
vacuum conditions for the representative choices $\tan\beta = 3$ and 10. 
As discussed in~\cite{EFGO}, lower values of $\tan\beta$ are almost
entirely excluded by LEP.
Our parameter scan was over the following ranges of parameters:
\begin{eqnarray}
0 & < m_0 < & 1000 \\
80 & < |\mu| < & 2000, \\
80 & < M_2 < & 1000, \\
0 & < m_A < & 1000, \\
-1000 & < A < & 1000.
\label{basic}
\end{eqnarray}
The main scan, which covers $m_0, \mu$ and $M_2 > 100$ GeV and $m_A >
300$ GeV, was supplemented with smaller but significant subscans, to cover
the smaller values of these four parameters as described below. The
values of
$m_0$ we use are fixed at the unification scale 
$\sim 10^{16}$~GeV, while the
values of the remaining parameters, $\mu, M_2, m_A,$ and $A$ are
evaluated at the electroweak scale. The lower cut off on both $M_2$ and
$\mu$ is due to the lower limit on the chargino mass. As we indicated
above, we impose a lower limit $m_A > 300$~GeV for $\mu < 0 $ to avoid
problems with $b \rightarrow s \gamma$. However, it should be noted that
this restriction is quite
conservative as, even for $m_A = 350 $~GeV, there are regions included in
the above scan which are not allowed by $b \rightarrow s
\gamma$~\cite{EFGO}.
Similarly, even for $\mu>0$, where we impose no cut off on $m_A$, we
have incuded some points which should be excluded on the basis of 
$b \rightarrow s \gamma$.

\begin{table}[htb]
\caption{\it
Details of MSSM parameter scans, including the numbers
of points that survive the LEP constraints and have a relic density
in the favoured range.}
\begin{center}
\begin{tabular}{cccc}
\hline\hline
 scan & Total points &  survived LEP  & $0.1 \le \Omega h^2 \le 0.3$\\
\hline
$M_2,\mu, m_0 \ge 100, m_A \ge 300$  &  &       &\\
\hline
$\tan \beta = 3, \mu > 0$ & 30000 &      17817 & 1552 \\
$\tan \beta = 3, \mu < 0$ & 30000 &      17210& 901\\
$\tan \beta = 10, \mu > 0$ & 30000 &      26498 & 2588\\
$\tan \beta = 10, \mu < 0$ & 30000 &     26507 & 2337\\
\hline
$100 \ge M_2,\mu \ge 80$, $m_0 \ge 100,  m_A \ge 300$  &  &       &\\
\hline
$\tan \beta = 3, \mu > 0$ & 20000 &      75 & 0 \\
$\tan \beta = 3, \mu < 0$ & 20000 &      4410& 30\\
$\tan \beta = 10, \mu > 0$ & 20000 &      1632 & 14\\
$\tan \beta = 10, \mu < 0$ & 20000 &     4480 & 58\\
\hline
$M_2,\mu \ge 80$, $m_0 \le 100,  m_A \ge 300$  &  &       &\\
\hline
$\tan \beta = 3, \mu > 0$ & 20000 &      2669 & 663 \\
$\tan \beta = 3, \mu < 0$ & 20000 &      2247&  487\\
$\tan \beta = 10, \mu > 0$ & 20000 &      5394 & 2436\\
$\tan \beta = 10, \mu < 0$ & 20000 &     5140 & 2377\\
\hline
$M_2,\mu \ge 80$, $m_0 \ge 0,  m_A \le 300$   &  &       &\\
\hline
$\tan \beta = 3, \mu > 0$ & 20000 &   2208    & 164 \\
$\tan \beta = 10, \mu > 0$ & 20000 &  12096     &  1170\\
\hline\hline
\end{tabular}
\end{center}
\label{tab:data}
\end{table}

\begin{figure}[htb]
\epsfig{figure=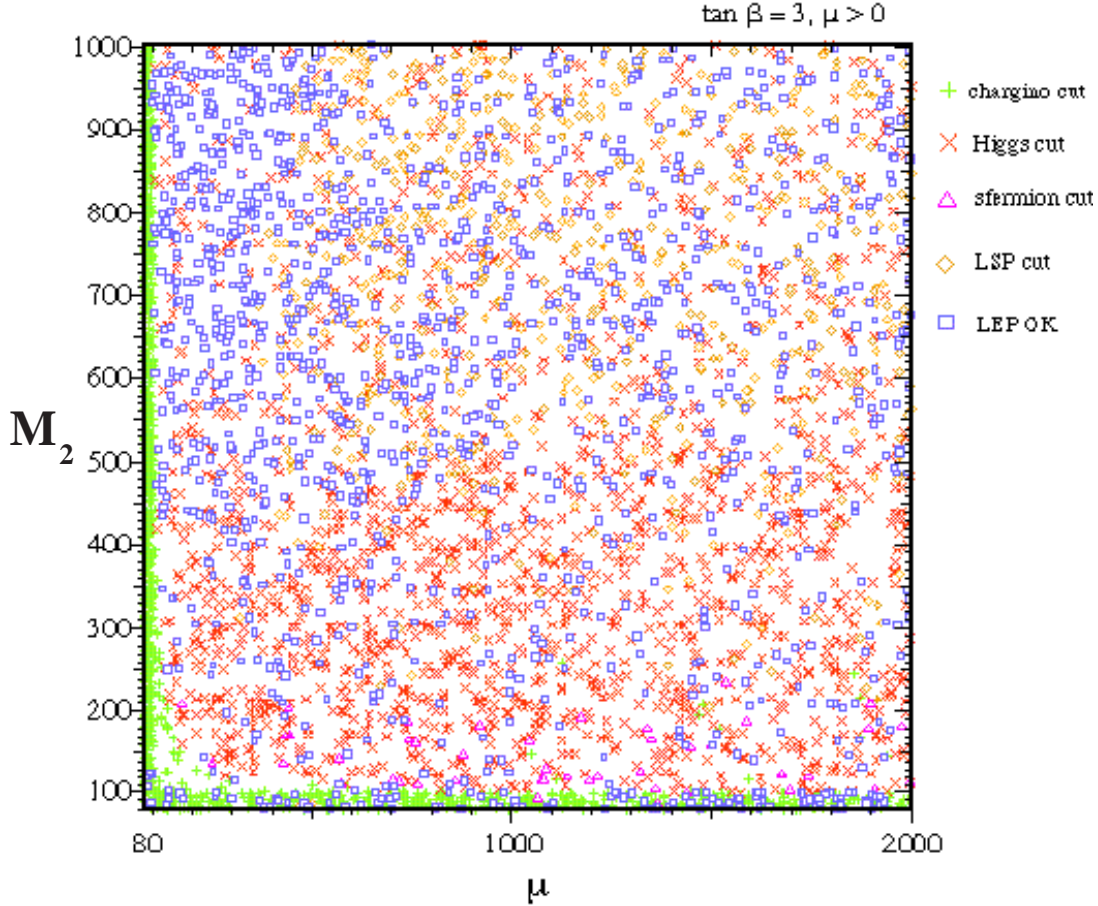,height=20cm}
\vskip -8cm
\caption{\it Results of the scan of MSSM parameter space 
for $\tan\beta = 3$ and $\mu > 0$ summarized in
Table~1, illustrating the impacts of the various LEP constraints. 
We denote by (green) pluses the points that fail the chargino
cut: $m_{\chi^\pm} \ge 101$~GeV, by (red) crosses the 
remaining points that survive the chargino cut but fail the
Higgs cut: $m_{H_2} \ge 104$~GeV, by (violet) triangles the 
points surviving the previous cuts that fail the sfermion cut: 
$m_{\tilde f} \ge 92$~GeV, and by (golden) diamonds
the points surviving the previous cuts that
do not have the lightest neutralino as the LSP. The (blue) squares 
denote scan points that survive all these LEP cuts.} 
\label{fig:LEPscan}
\end{figure}

As can be seen in the Table, the overall scan was divided into
three (four) specific regions for each value of $\tan \beta$ and
$\mu$ negative (positive), each with the number of
points listed. The subscans with lower thresholds were designed to scour
carefully the regions of MSSM parameter space close to the LEP
exclusions, with the aim of ensuring that we sampled points close to
their boundaries. For each subscan, we show the number of points
which survive all the LEP experimental constraints discussed
above, and we see that lower fractions of the low-threshold
subscans survive them, in particular because they tend to
yield excluded values of the chargino mass. Fig.~\ref{fig:LEPscan}
provides some insight into the impacts of the different LEP
constraints for the case $\tan \beta = 3$ and $\mu > 0$.
We plot in Fig.~\ref{fig:LEPscan} the points scanned in the $M_2 - \mu$
parameter plane.  In making this scatter plot, we show a randomly chosen
subset of 5000 of the 90000 points sampled~\footnote{We have checked that
there is no
qualitative difference between this plot and the much denser plot with
all points shown.}, since it is much easier to pick out the relevant
physical effects of the cuts in such a subset of points, the full plot
being extremely dense.  

We see that the chargino cut removes points at low values of
$\mu$ and $M_2$, denoted by (green) pluses, that the Higgs cut then
removes many more points with low $M_2$, denoted by (red) crosses, that
the sfermion cut removes still
more points with low $M_2$, denoted by (violet) triangles (this occurs at
high $A$ and/or $\mu$ when there is a sizeable off-diagonal component in
the sfermion mass matrix), and that the LSP cut tends to remove points at
higher
$M_2$ denoted by (golden) diamonds. The surviving (blue) squares are
spread over the
$\mu, M_2$ plane, except for small values. Note that some points may fail
to survive more than one of the above cuts. These are only denoted by the
first cut tested and failed in the order listed above.  The scans for the
opposite sign of
$\mu$ and for
$\tan\beta = 10$ exhibit similar features,
and are omitted here. The only noticeable difference 
when $\mu > 0$ is that not so
many were points eliminated by the Higgs cut at large values of $M_2$ for
$\mu < 0$, because we imposed the limit $m_A > 300$~GeV:
for $\mu > 0$,
many more points were run with low $m_A$ and hence low $m_{H_2}$. 
Also, for $\tan \beta = 10$, more points survive at low $\mu$ and/or $M_2$
due to the relaxed contraint on the Higgs mass.

\begin{figure}[htb]
\vskip -.5cm
\epsfig{figure=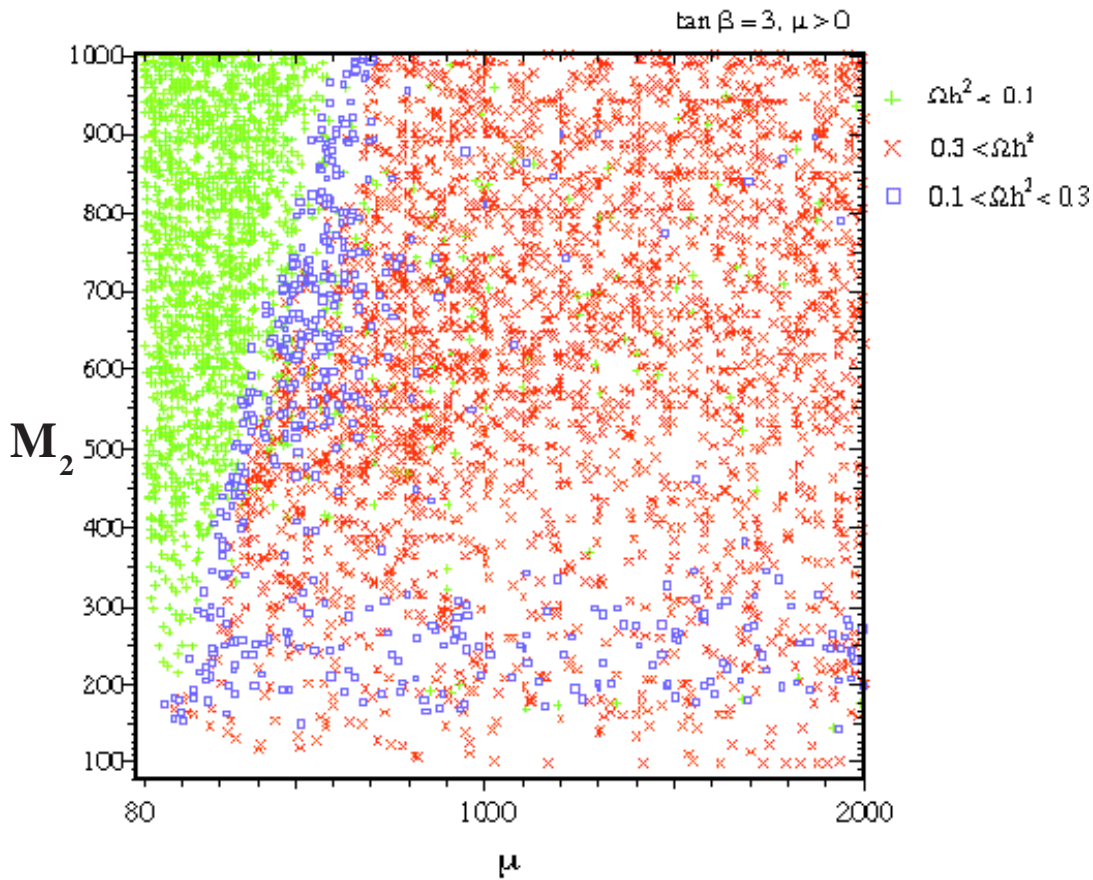,height=20cm}
\vskip -8cm
\caption{\it Results of the scan of MSSM parameter space
for $\tan\beta = 3$ and $\mu > 0$ summarized in
Table~1, illustrating the impact of the cosmological 
relic density constraint on the points that points that
survived the LEP constraints illustrated in Fig.~\ref{fig:LEPscan}.
We denote by (green) pluses the points that have too small
a relic density: $\Omega_\chi h^2 < 0.1$,
by (red) crosses the
points that have too high a relic density: $\Omega_\chi h^2 > 0.3$,
and by (blue) squares the good points for which
$0.1 \le \Omega_\chi h^2 \le 0.3$.}
\label{fig:cosmoscan}
\end{figure}

The last column of the Table shows how many of the points that
survive the LEP constraints and
have relic densities in the cosmologically preferred range
$0.1 \le \Omega_\chi h^2 \le 0.3$. It is apparent that
most of the preferred points emerge from the first scan
with $M_2,\mu > 100$, as the lower values which were explored thoroughly
in the second subscan generally failed the chargino cut. More
details of the scan over cosmological relic densities 
for $\tan \beta = 3$ and $\mu > 0$ are shown in
Fig.~\ref{fig:cosmoscan}. As in Fig.~\ref{fig:LEPscan}, we show only 
a randomly selected subset of 5000 points out of the total of
approximately 22000 points which survived the LEP cuts.
We see that, among the points that survived the previous LEP
constraints, those with a small ratio of $\mu / M_2$ generally
have too small a relic density, denoted by (green) pluses, as a result of
over-efficient
$\chi - \chi' - \chi^\pm$ coannihilation, whereas points with
$\mu / M_2 \sim 1$~to~5 tend to have too large a relic density,
denoted by (red) crosses,
particularly if $\mu$ and $M_2$ are individually large. The
points with a relic density in the preferred range
$0.1 \le \Omega_\chi h^2 \le 0.3$, denoted by (blue) squares, tend to
accumulate around
$\mu / M_2 \sim 1/2$ or low $M_2$. The former points are in
the transition region between over-efficient
$\chi - \chi' - \chi^\pm$ coannihilation and under-efficient
annihilation at large $\mu$ and $M_2$, whereas the latter are in
the region of low $M_2$ where careful implementation of the LEP
constraints is essential. However, it is apparent from
Fig.~\ref{fig:cosmoscan} that there are exceptions to these
general trends. We do not discuss them in detail, but remark that
we have made an attempt to understand at least those exceptions that
lead to `unusual' elastic scattering cross sections.

\section{Elastic Scattering Cross Sections}

\begin{figure}[htb]
\vskip -3cm
\hspace*{-.35in}
\begin{minipage}{8in}
\epsfig{figure=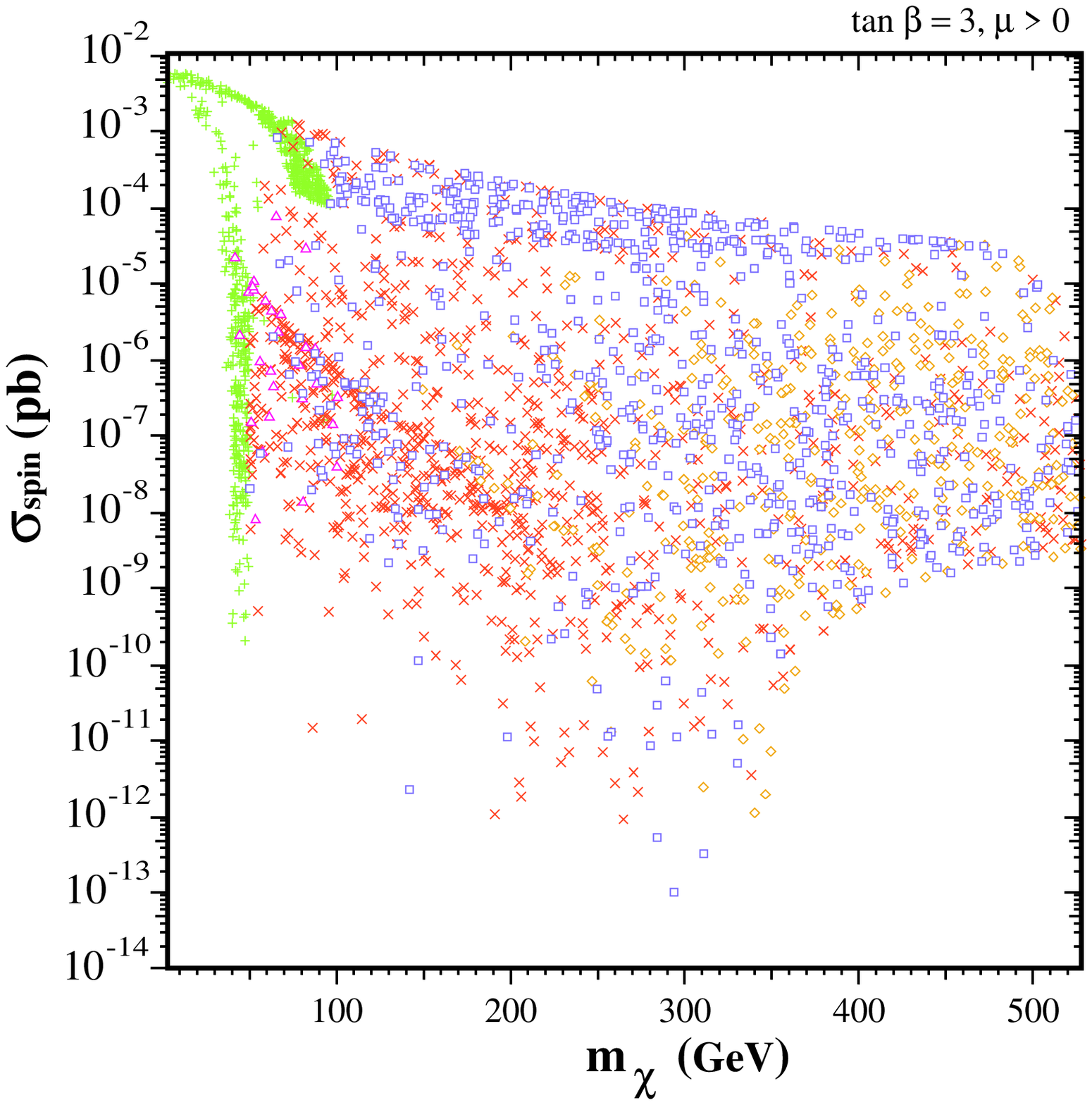,height=5.in}
\hspace*{-.65in}\epsfig{file=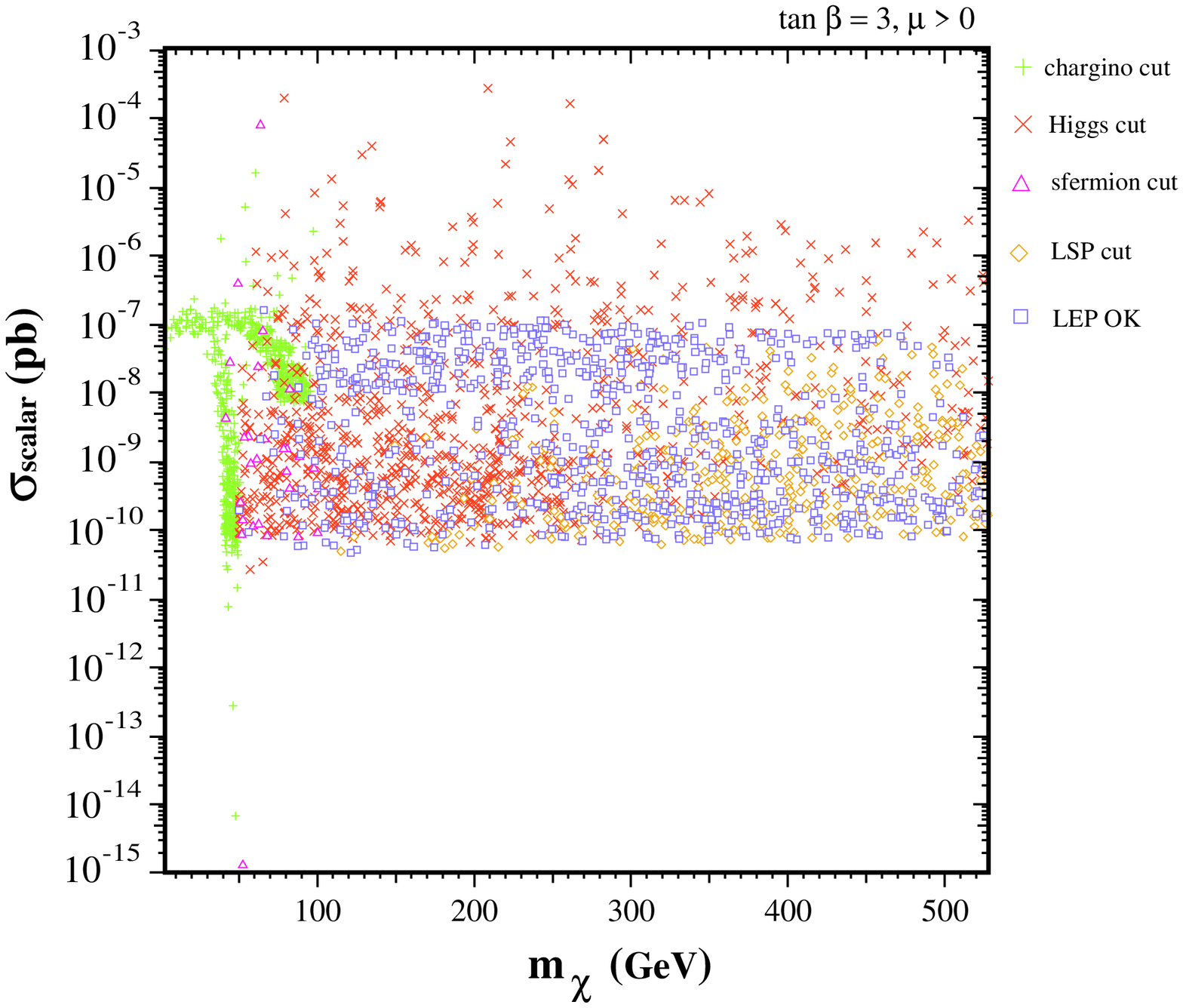,height=5.in} \hfill
\end{minipage}
\vskip -1.5cm
\caption{\it Scatter plots of (a) the spin-dependent and (b) the
spin-independent elastic scattering cross sections for
$\tan\beta = 3, \mu > 0$ for a representative subsample of
3000 points, illustrating the impacts of the
LEP constraints. The (green) plus signs denote points that fail the
chargino mass constraint, which usually have small $m_\chi$ and
sometimes large cross sections. The (red) crosses denote 
surviving points that fail the Higgs mass constraint, some of which
have large spin-independent cross sections. The (violet) triangles denote
surviving points that fail our (approximate) sfermion mass cut. The
(golden)
diamonds denote points where $\chi$ is not the LSP, and the (blue)
squares denote points that survive all the LEP cuts.}
\label{fig:sigmaLEP}
\end{figure}

We now discuss the values of the elastic scattering cross sections that
are attainable, bearing in mind the LEP and cosmological relic density
constraints. Fig.~\ref{fig:sigmaLEP} illustrates the allowed ranges
of elastic scattering cross sections for the points included in
our scan for the particular case $\tan\beta = 3, \mu > 0$, as it was
described in the previous Section.
Plotted is a subset of 3000 of the 90000 points scanned, indicating
which points survive all the LEP cuts, and which other points fail which
LEP cut.
We find similar results for
$\tan\beta = 10$ and/or the opposite sign of $\mu$, with the exception
that when $\mu<0$ we find some points trickling down below the apparent
boundary at $ \sim 10^{-10}$ pb in Fig.~\ref{fig:sigmaLEP}(b), because
of cancellations similar to those discussed in~\cite{EFlO1}.

We note, in particular, that the LEP chargino and Higgs cuts
remove many points with low $m_\chi$ and/or large elastic scattering
cross sections. The sfermion mass cut is less important.
The constraint that $\chi$ be the LSP removes quite a large number of
points, populated more or less evenly in these cross section plots.
The
somewhat sparse set of points with very small cross sections give
some measure of how low the cross section may fall in some special cases.
These reflect instances where particular cancellations take place,
examples of which were discussed in~\cite{EFlO1}, and should not be
regarded as generic. The lower boundary of the densely occupied region in
Fig.~\ref{fig:sigmaLEP} offers an answer to the question how low
the elastic scattering cross sections may reasonably fall, roughly
$\sigma \sim 10^{-9}$~pb for the spin-dependent cross section and
$\sim 10^{-10}$~pb for the spin-independent cross section.

We would like to draw particular attention to the spin-independent 
cross-section shown in Fig.~\ref{fig:sigmaLEP}(b). Notice that there
are parameter choices with very large scattering cross sections.
In this random selection, the cross-section may be as high as a few
$\times 10^{-4}$ pb, and could even be larger than that claimed by DAMA.
Indeed, in the full set of 90000 points scanned, there are even a few
points which surpass $10^{-3}$ pb.  However, {\em all} of these points 
have been excluded by LEP (primarily by the Higgs mass cut). The
largest surviving cross section is slightly over $10^{-7}$ pb,  in both
the randomly selected subset and the full scan. For $\mu<0$, the upper
boundary in the scalar cross section is about an order of magnitude lower,
as was the case in the model with universal Higgs masses~\cite{EFlO1}.
Note also that, for $\mu < 0$, the limit $m_A >  300$ GeV we impose
removes the points with large cross sections (in this case with
$\sigma_{\rm scalar} \ga 10^{-8}$ pb).

\begin{figure}[h]
\vskip -5cm
\hspace*{-.60in}
\begin{minipage}{8in}
\epsfig{figure=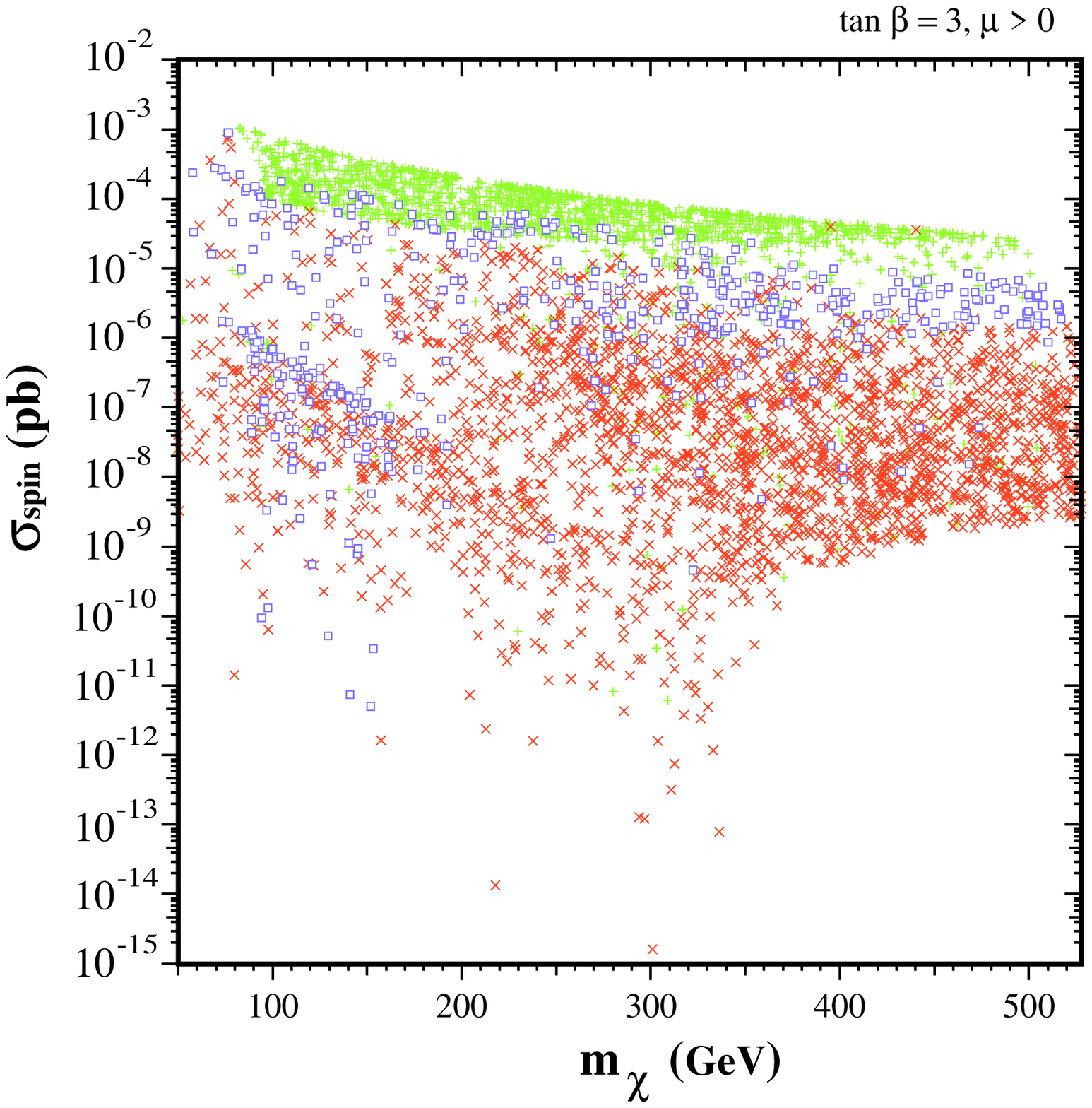,height=5.5in}
\hspace*{-.70in}\epsfig{file=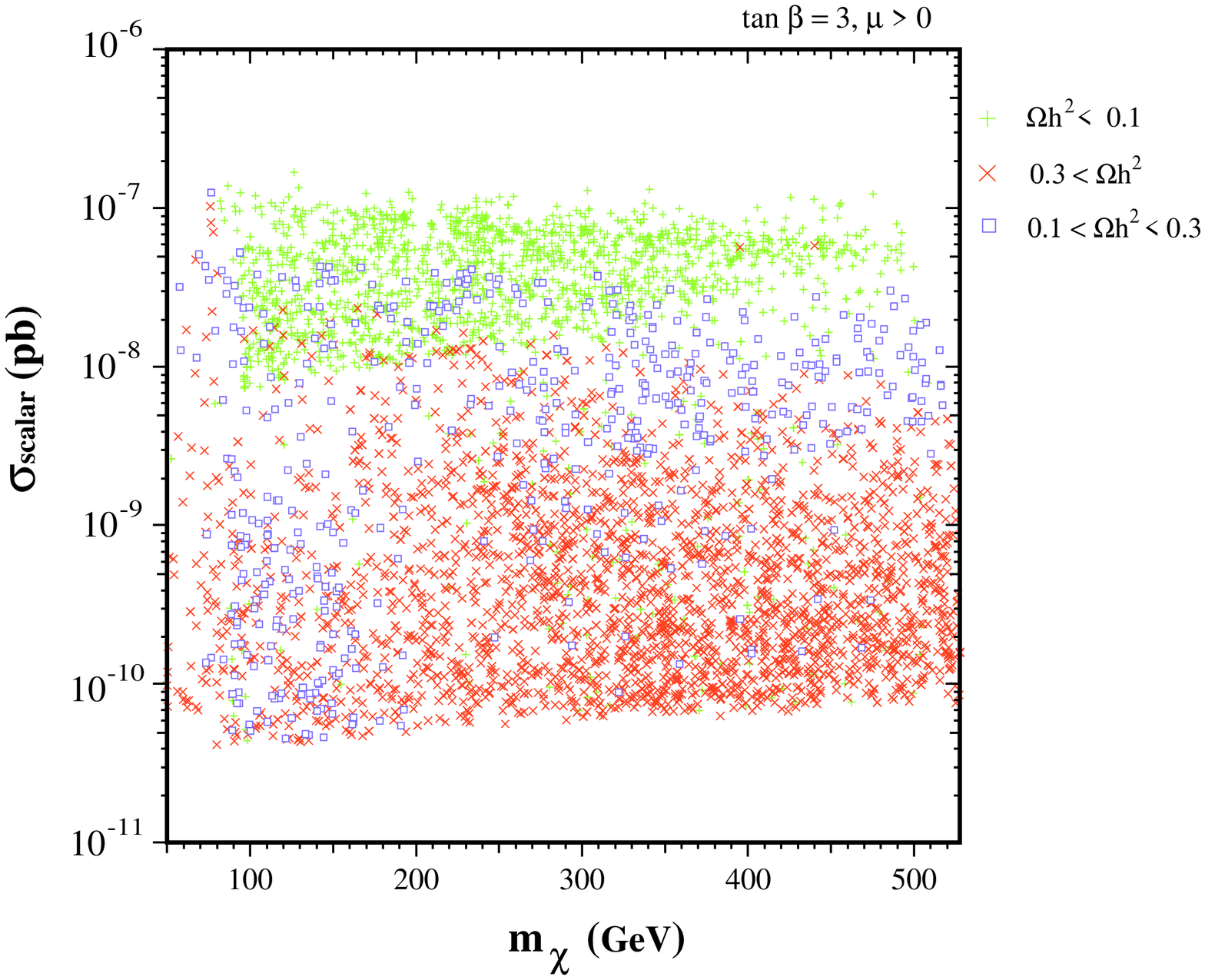,height=5.5in}
\hfill
\end{minipage}
\vskip -5.5cm
\hspace*{-.60in}
\begin{minipage}{8in}
\epsfig{figure=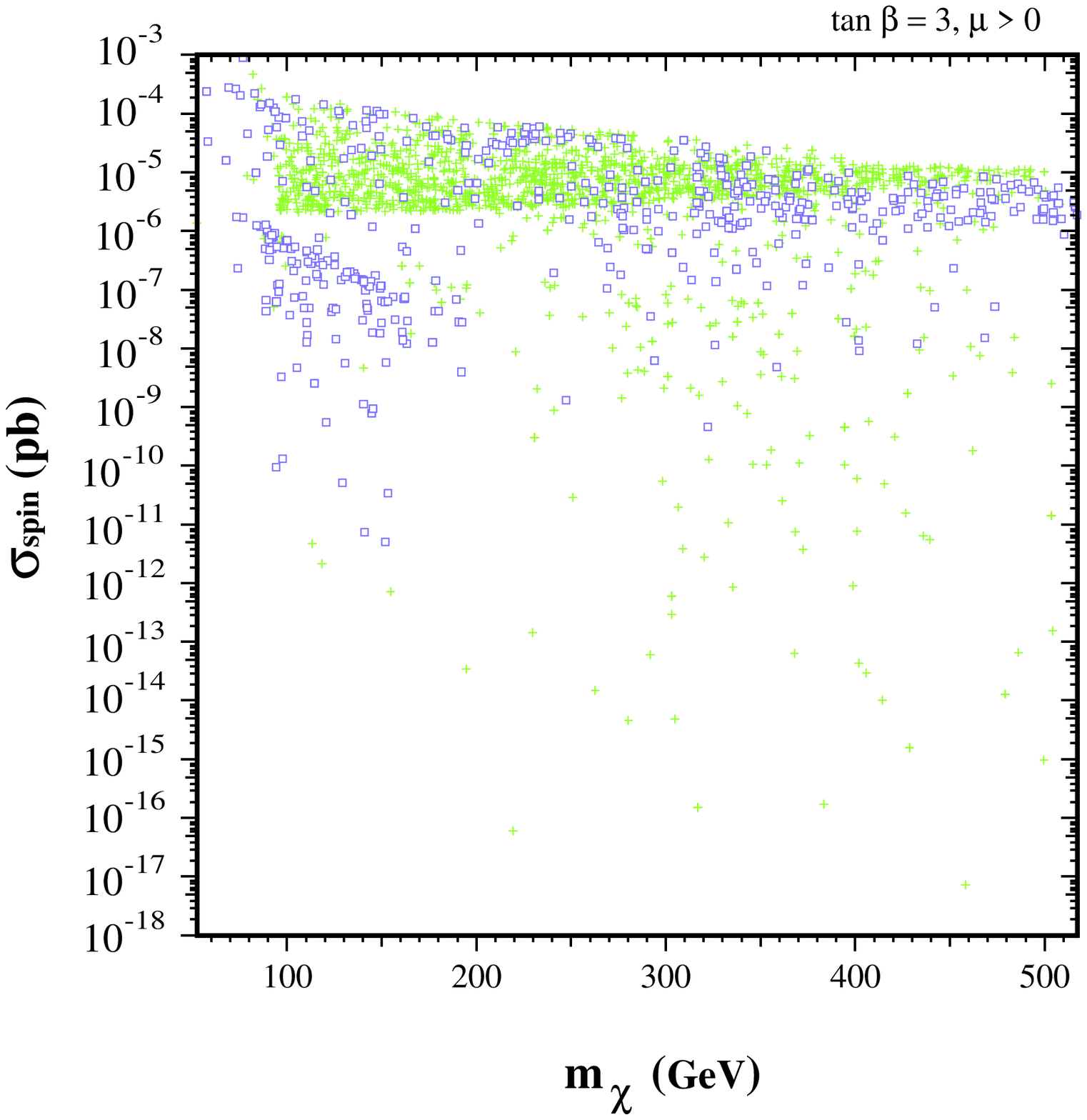,height=5.5in}
\hspace*{-.70in}\epsfig{file=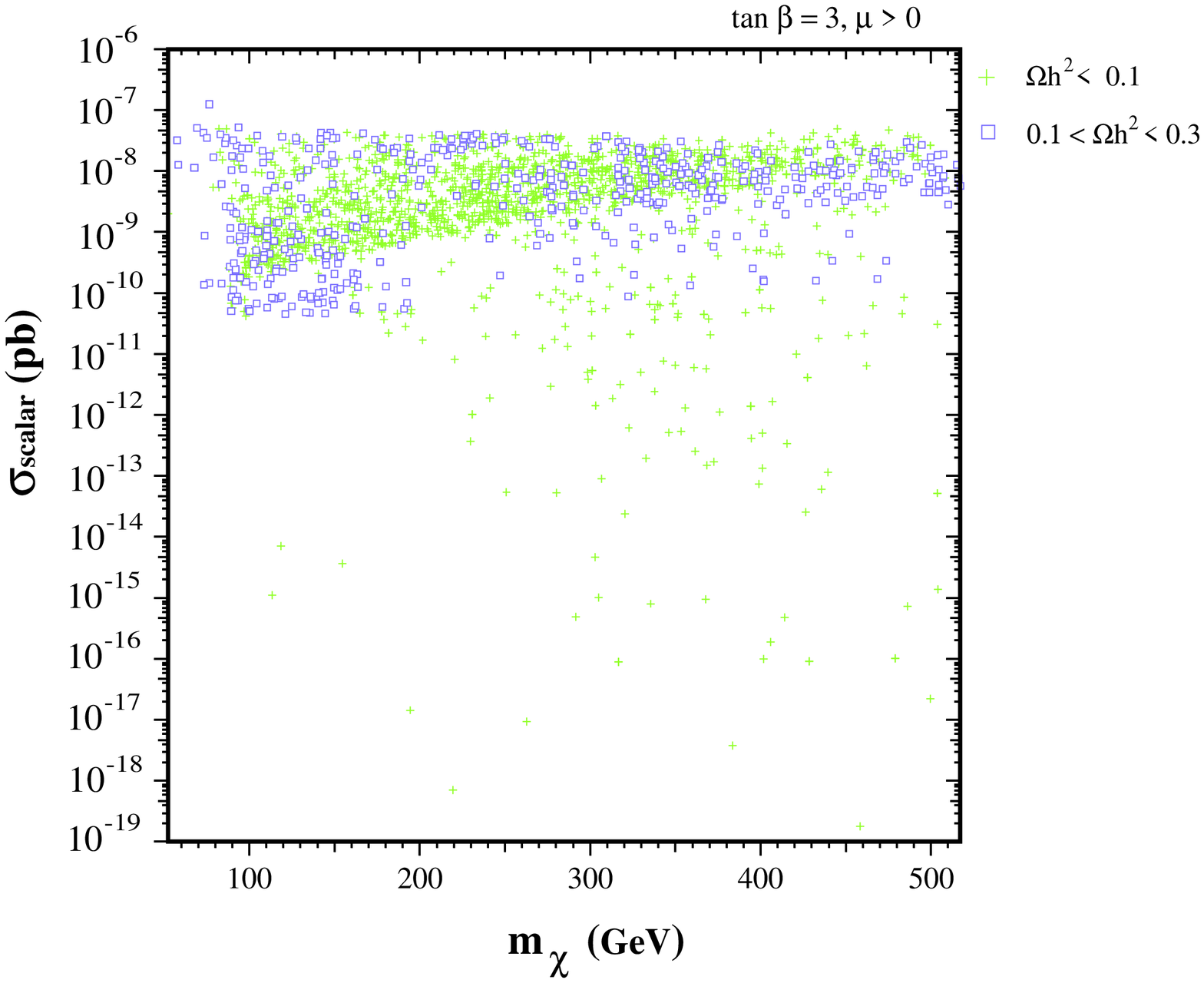,height=5.5in}
\end{minipage}
\vskip -1.5cm
\caption{\it Scatter plots (a,c) of the spin-dependent and (b,d)
of the spin-independent elastic scattering cross sections for
$\tan\beta = 3, \mu > 0$, after implementing the LEP constraints,
exhibiting the impacts of the 
cosmological relic density constraints. The (green) pluses
have $\Omega_\chi h^2 < 0.1$, the favoured (blue) boxes
have $0.1 \le \Omega_\chi h^2 \le 0.3$, and the (red) crosses have
$\Omega_\chi h^2 > 0.3$. Note in panels (c) and (d) the impacts of
removing the over-dense points, which tend to have lower cross sections,
and rescaling the under-dense points as in (\ref{halofraction}),
suppressing some high cross section points.}
\label{fig:sigmacosmo}
\end{figure}

The next step is to implement the cosmological relic density
constraints. We show in Fig.~\ref{fig:sigmacosmo} the cross sections
obtained for a representative subsample of points with $\tan\beta = 3,
\mu > 0$ that survive the LEP cuts, sorted according to the calculated
values of $\Omega_\chi h^2$.
Spin-dependent cross sections are plotted in panels (a) and (c), and
spin-independent cross sections are plotted in panels (b) and (d).
We include in panels (a) and (b) the cross sections calculated
for unrealistic models with $\Omega_\chi h^2 > 0.3$, and
without making any rescaling correction for points with
$\Omega_\chi h^2 < 0.1$. The over-dense points with $\Omega_\chi h^2 > 
0.3$, denoted by (red) crosses, have been removed in panels (c)
and (d), and the cross sections for
under-dense points with $\Omega_\chi h^2 < 0.1$, denoted by (green)
pluses, have been rescaled by
the appropriate halo density fraction (\ref{halofraction}). As could
be expected, the over-dense points tend to have smaller cross sections,
and the under-dense points larger cross sections before applying the
rescaling correction. After rescaling, the under-dense points yield
cross sections in the range found for the favoured points with
$0.1 \le \Omega_\chi h^2 \le 0.3$, denoted by (blue) boxes.
For $\tan \beta = 10$ and $\mu > 0$, the scalar cross section is about an
order of magnitude higher for points which survive all cuts. 
Relative to the cases with $\mu > 0$, the $\mu <0$ cases have a scalar
cross section which is 1-2 orders of magnitude smaller. 

A comparison with Fig.~\ref{fig:cosmoscan} shows that the largest
cross sections displayed in Figs.~\ref{fig:sigmacosmo}(a,b), are almost
all for
Higgsino-like states whose elastic cross section is mediated by
$Z$ exchange.  These are cosmologically under-dense, due to a combination
of
large annihilation and coannihilation cross sections. The cosmologically
over-dense regions with relatively low elastic cross sections are
mainly for
gaugino-like states, and are for the most part more massive than 300 GeV,
which is the oft-quoted upper bound on the bino mass in the MSSM
\cite{up}. 

\begin{figure}[htb]
\vskip -3cm
\hspace*{-.60in}
\begin{minipage}{8in}
\epsfig{figure=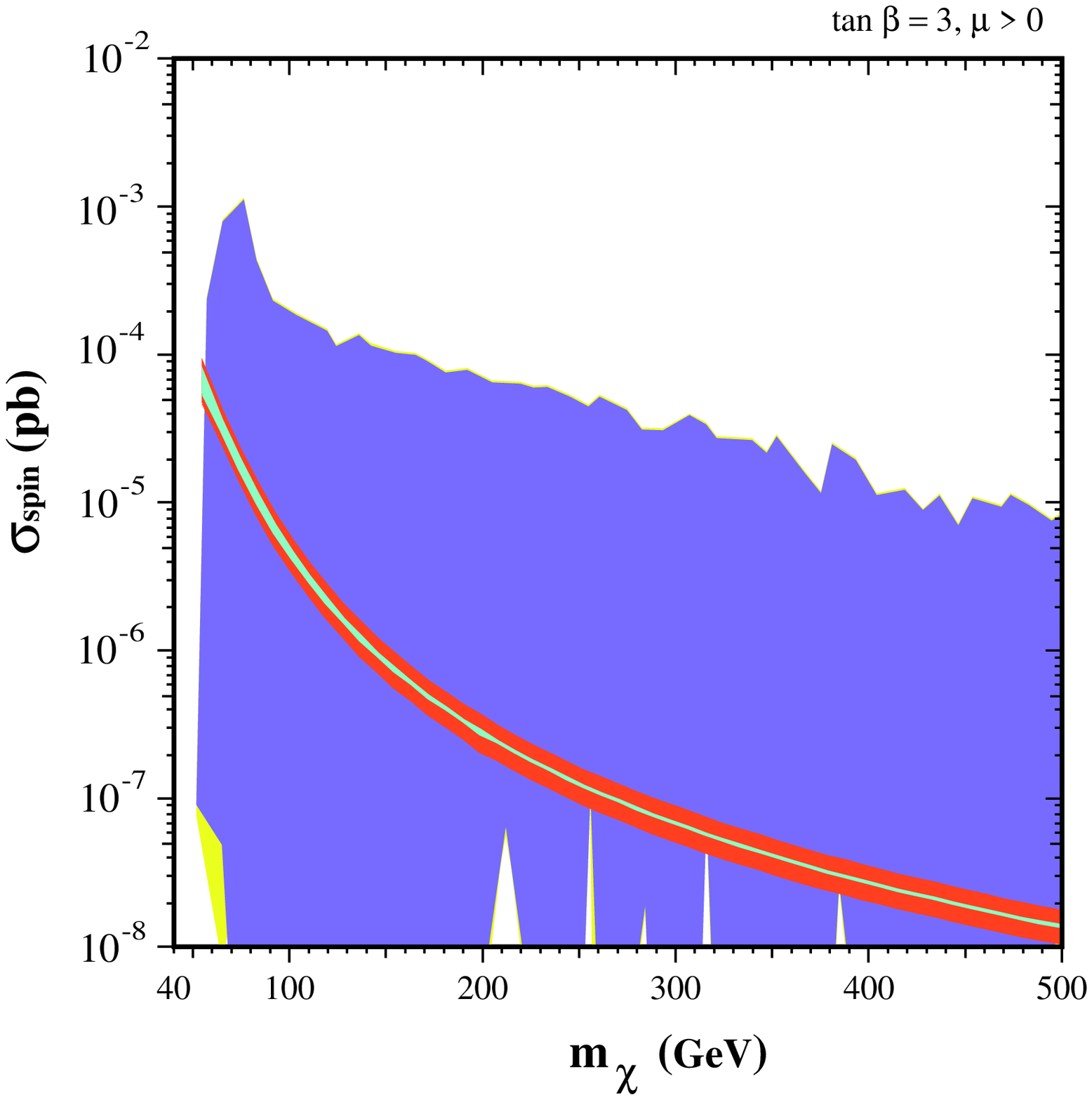,height=5.5in}
\hspace*{-.70in}\epsfig{file=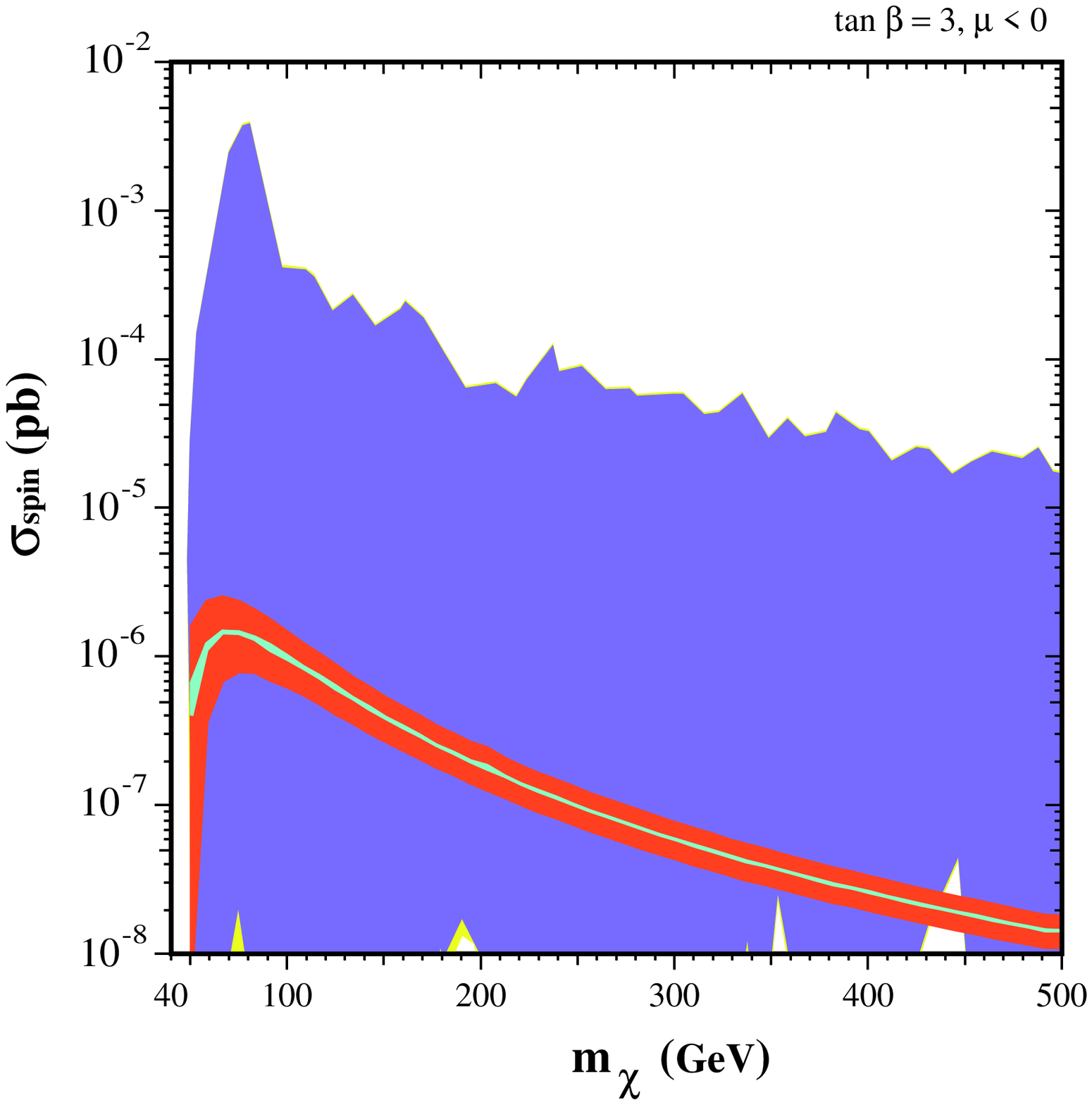,height=5.5in}
\hfill
\end{minipage}
\vskip -5cm
\hspace*{-.60in}
\begin{minipage}{8in}
\epsfig{figure=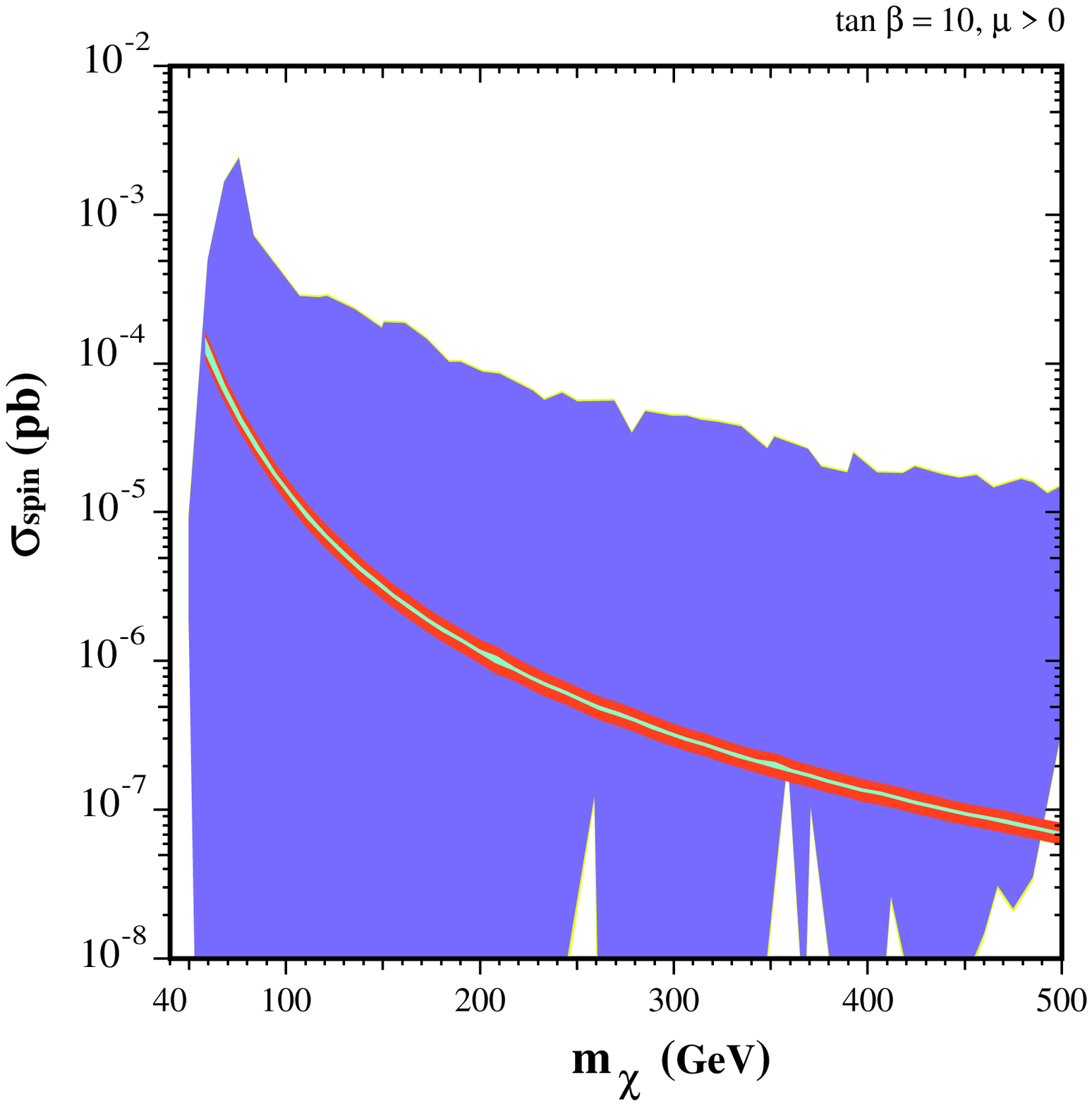,height=5.5in}
\hspace*{-.70in}\epsfig{file=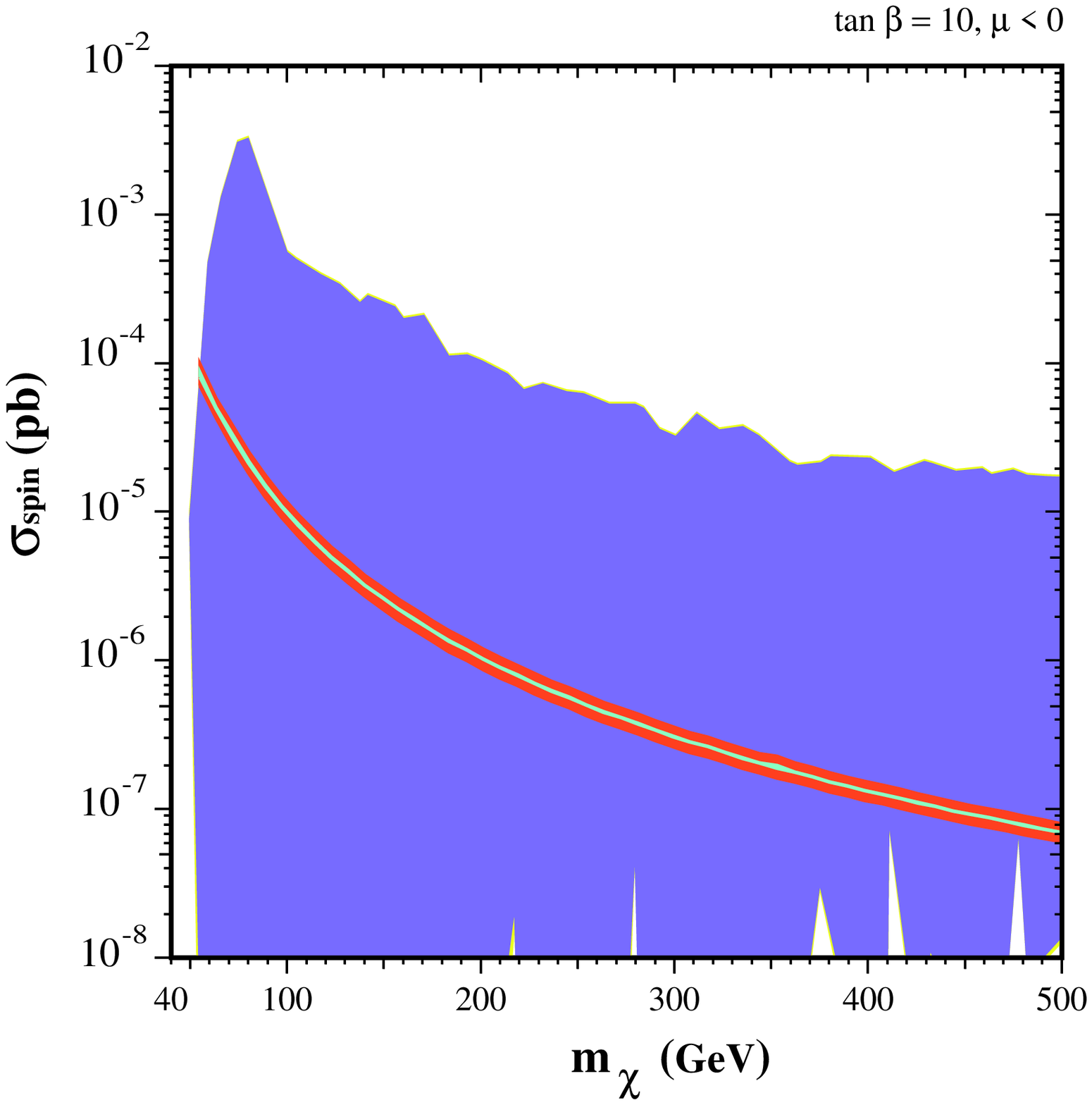,height=5.5in}
\end{minipage}
\vskip -1.5cm
\caption{\it 
Allowed ranges of the spin-dependent elastic neutralino-proton
cross section for (a) $\tan\beta = 3$ and $\mu > 0$, (b)
$\tan\beta = 3$
and $\mu < 0$, (c) $\tan\beta = 10$ and $\mu > 0$ and (d)
$\tan\beta = 10$
and $\mu < 0$. The main (blue) shaded
regions summarize the envelopes of possible values found in our scan,
for points respecting the LEP constraints, discarding points with
$\Omega_\chi h^2 > 0.3$, and rescaling points with $\Omega_\chi h^2 <
0.1$ according to (\ref{halofraction}). The small light (yellow) shaded
extensions of this region reflect the hadronic matrix element
uncertainties discussed in Section 2. The concave
(red and turquoise) strips are those
found previously assuming universal Higgs scalar masses
(UHM)~\cite{EFlO1}.}
\label{fig:spin}
\end{figure}

Our resulting predictions for the spin-dependent elastic neutralino-proton
cross section for $\tan\beta = 3$ and $\mu > 0$, after
taking into account the LEP and cosmological constraints, are shown in
Fig.~\ref{fig:spin}(a), where a comparison with the UHM case is also
made~\footnote{In contrast to~\cite{EFlO1}, here we have
taken into account the updated LEP constraints.}.
The raggedness of the upper and lower boundaries of the dark (blue) shaded
allowed
region reflect the coarseness of our parameter scan, and the relatively
low density of parameter choices that yield cross sections close to
these boundaries. 
We see that, at low $m_\chi$ close to the LEP limit,
the spin-dependent cross section may be as much as an order of magnitude
greater than
in the UHM case considered previously~\cite{EFlO1}, shown by the concave
(red and turquoise) strip. However, even for low
$m_\chi$, the attainable range is far below the present experimental
sensitivity, which is to $\sigma_{spin} \sim 1$~pb, and could be many
orders of magnitude lower. As $m_\chi$ increases, the maximum allowed
value of $\sigma_{spin}$ decreases, though not as rapidly as in the
previous UHM case~\cite{EFlO1}. The hadronic uncertainties are basically
negligible for this spin-dependent cross section, as seen from the
light (yellow) shading.
Turning now to the option $\tan\beta = 3$ and $\mu < 0$
shown in Fig.~\ref{fig:spin}(b), we see that
the allowed range of the spin-dependent cross section is similar to
that in the $\tan\beta = 3, \mu > 0$ option. This is in contrast to
the situation in the UHM~\cite{EFlO1}, where the spin-dependent
cross section at low $m_\chi$ is much smaller for $\mu < 0$ than for $\mu
> 0$. However, the cross section is still three or more orders of
magnitude away from the present experimental upper limit.
In the option $\tan\beta = 10$ and $\mu > 0$ shown in
Fig.~\ref{fig:spin}(c), we see that the
attainable range of the spin-dependent cross section is again similar
to the previous option. This again contrasts with the UHM case, where
the narrow allowed band for large $m_\chi \sim 500$~GeV
was somewhat higher than for the option $\tan\beta = 3$ and $\mu > 0$.
As shown in Fig.~\ref{fig:spin}(d), our results for $\tan\beta = 10$ and
$\mu < 0$ are very similar to those for $\mu > 0$.

\begin{figure}[htb]
\vskip -3cm
\hspace*{-.60in}
\begin{minipage}{8in}
\epsfig{figure=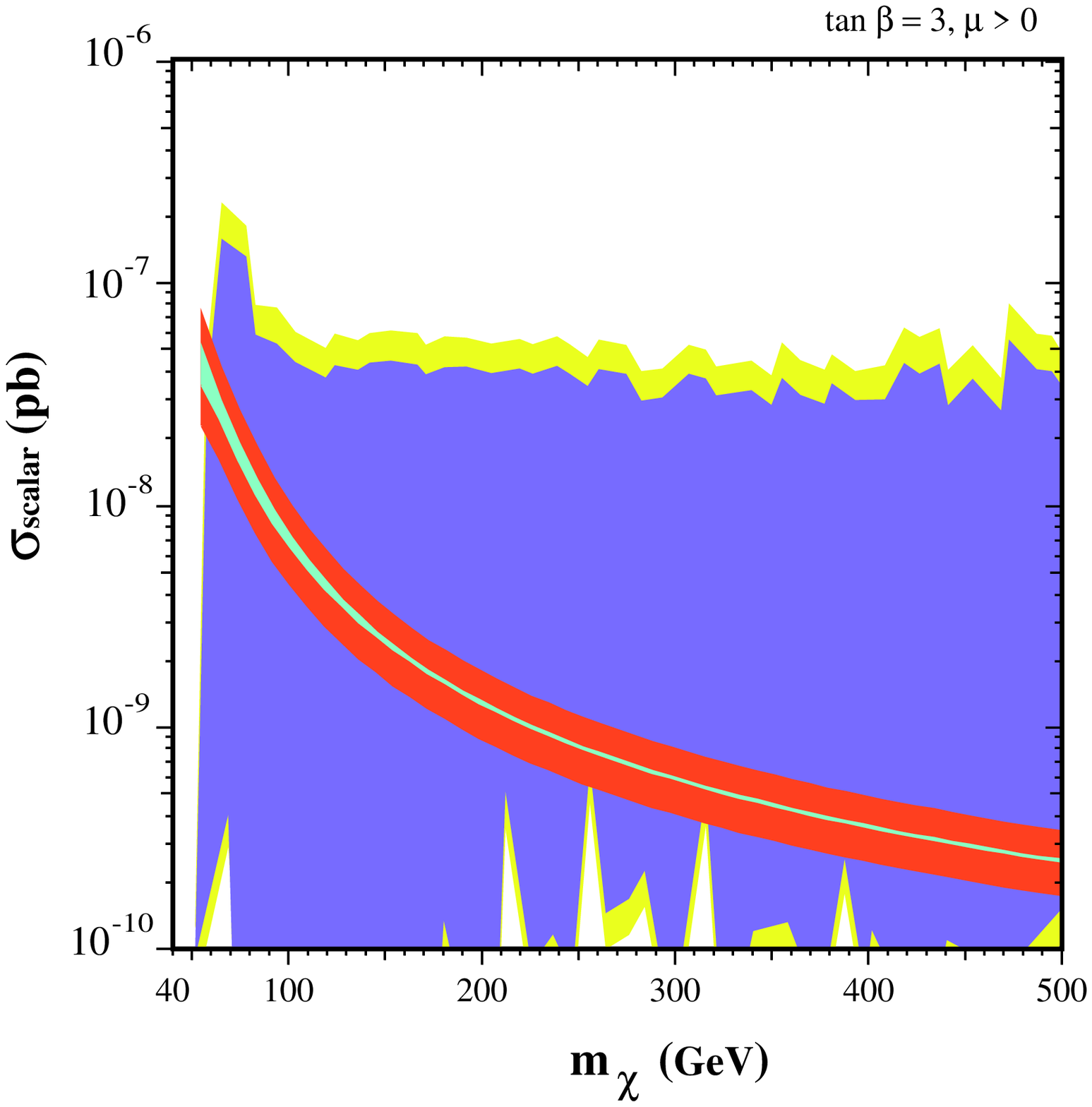,height=5.5in}
\hspace*{-.70in}\epsfig{file=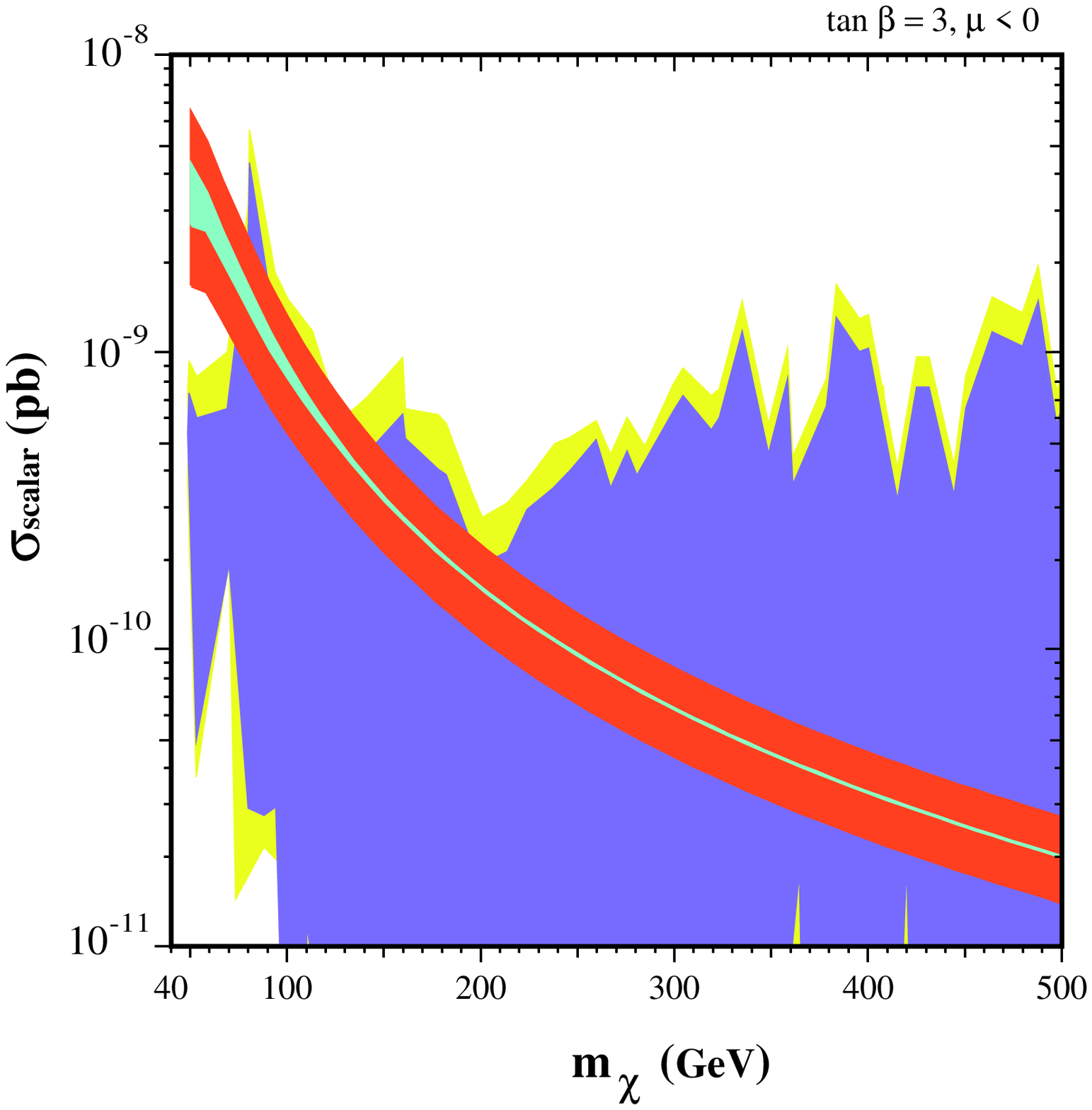,height=5.5in}
\hfill
\end{minipage}
\vskip -5cm
\hspace*{-.60in}
\begin{minipage}{8in}
\epsfig{figure=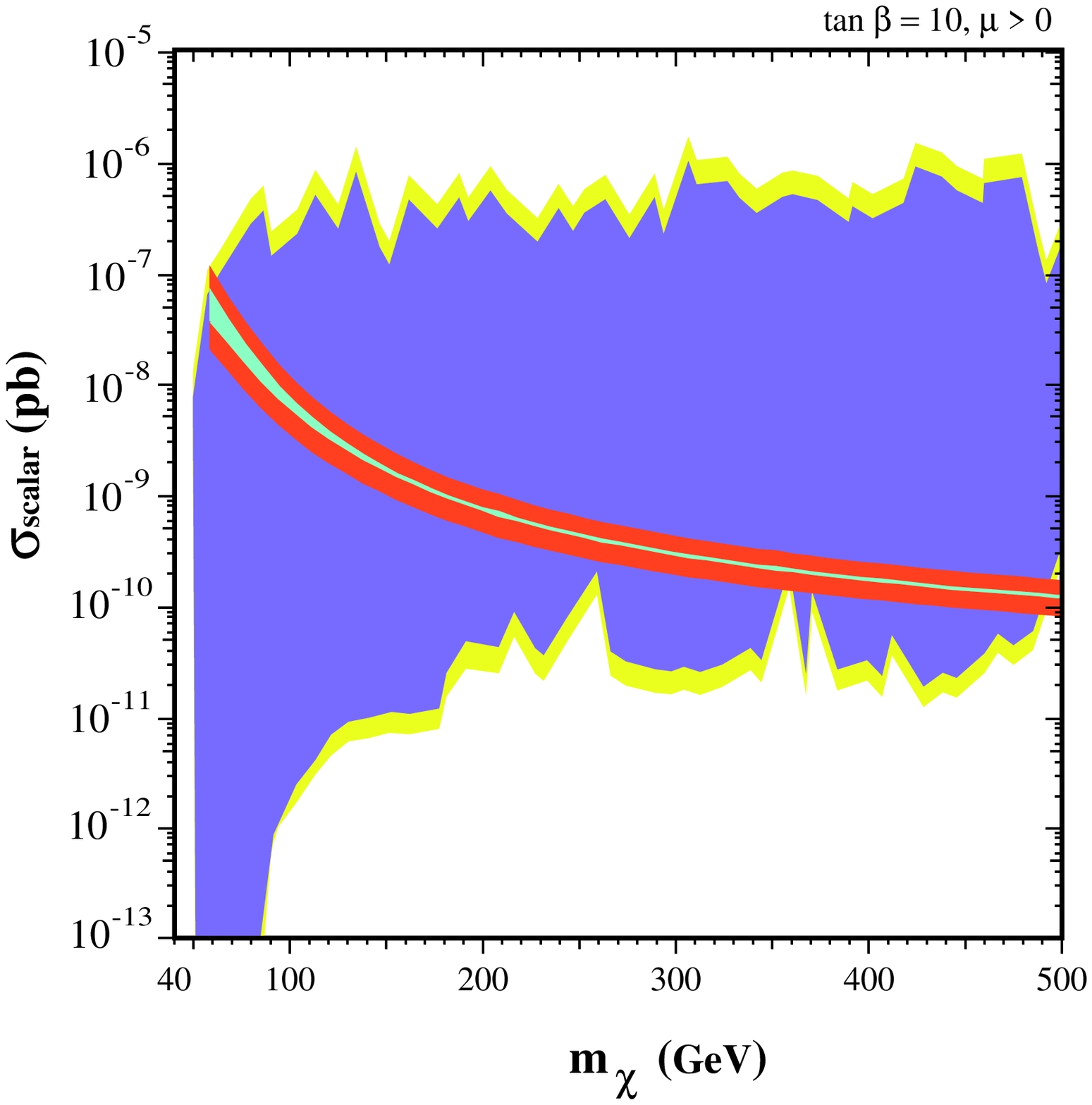,height=5.5in}
\hspace*{-.70in}\epsfig{file=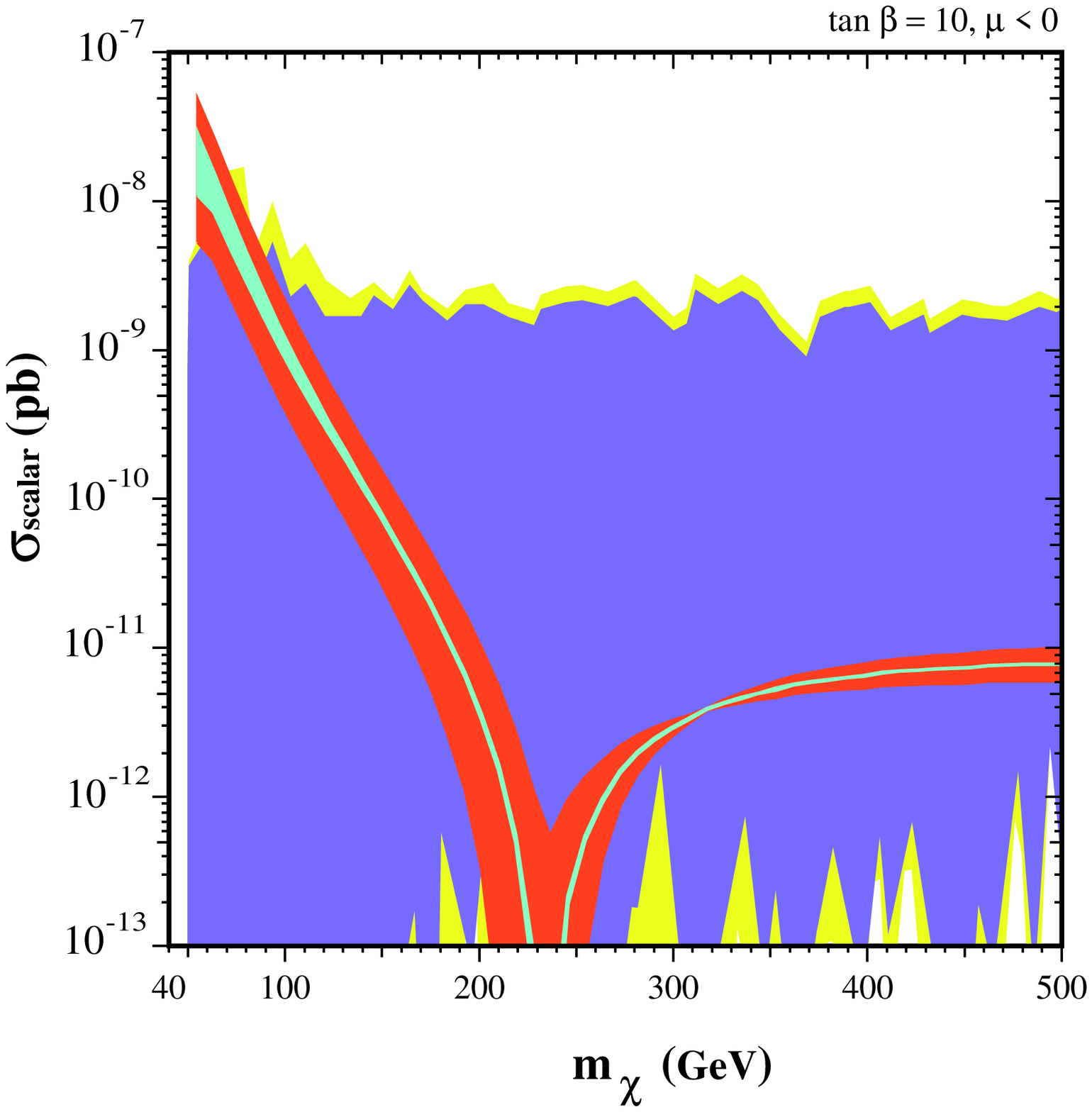,height=5.5in}
\end{minipage}
\vskip -1.5cm
\caption{\it 
Allowed ranges of the spin-independent elastic neutralino-proton
cross section for (a) $\tan\beta = 3$ and $\mu > 0$, (b) $\tan\beta = 3$
and $\mu < 0$, (c) $\tan\beta = 10$ and $\mu > 0$ and (d) $\tan\beta = 10$ 
and $\mu < 0$. The main (blue) shaded
regions summarize the envelopes of possible values found in our scan,
for points respecting the LEP constraints, discarding points with
$\Omega_\chi h^2 > 0.3$, and rescaling points with $\Omega_\chi h^2 <
0.1$ according to (\ref{halofraction}). The small light (yellow) shaded
extensions of this region reflect the hadronic matrix element
uncertainties discussed in Section 2. The 
(red and turquoise) diagonal strips are the results
found assuming universal Higgs scalar masses (UHM)~\cite{EFlO1}.}
\label{fig:scalar}
\end{figure}

The analogous results for the spin-independent elastic
neutralino-proton cross section, after
taking into account the LEP and cosmological constraints, are shown in  
Fig.~\ref{fig:scalar}, where comparisons with the UHM case are also
made.
We see in panel Fig.~\ref{fig:scalar}(a) for $\tan \beta = 3$ and $\mu >
0$ a pattern
that is similar to the spin-dependent case.
For small $m_\chi$, the spin-independent
scalar cross section, shown by the dark (blue) shaded region, may be
somewhat higher than in the UHM case, shown by the (red and turquoise)
diagonal strip, whilst it could be much smaller. For large $m_\chi$,
the cross section may be rather larger than in the UHM case, but it is
always far below the present sensitivity.  The case
shown in panel (b) of
$\tan \beta = 3$ and $\mu < 0$ is somewhat different: the cross section
never gets to be significantly larger than the UHM value at small
$m_\chi$. The reason for the anomalous extension of the UHM band outside
the more general range is that the newer analysis reflected in the
(blue and yellow) shaded region incorporates updated LEP
constraints~\cite{EFGO}, that
are significantly stronger for small $\tan \beta$ and small $m_\chi$ than
those used in~\cite{EFlO1}.
This `anomaly' is absent in panel (c) for $\tan \beta = 10$ and $\mu < 0$,
which closely resembles panel (a), and also panel (d)  for $\tan \beta =
10$ and $\mu > 0$. We note in panel (d) a lesser reappearance of the
`anomalous' outdated UHM region at small $m_\chi$. The
dip in the (red and turquoise) UHM band for
$m_\chi \sim 230$~GeV in panel (d) reflects rather special
cancellations~\cite{EFlO1} that are absent in the more general case.
Overall, we note that the hadronic uncertainties, denoted by the light
(yellow) bands, are somewhat larger in the spin-independent case
than in the spin-dependent case.

\section{Summary and Prospects}

In this paper we have extended the analysis of~\cite{EFlO1} to
consider a more general sampling of supersymmetric models, relaxing
the UHM assumption we made previously. For each of two choices of
$\tan \beta$ and $\mu$ negative (positive), we have sampled 70000
(90000) sets
of MSSM parameters, 30000 in general scans and 20000 each in two (three)
special subscans over lower values of $M_2, \mu, m_0$ (and $m_A$). We
have implemented the current LEP constraints on MSSM
parameters~\cite{EFGO}, discussing in detail which scan points
survive which of these constraints. We have further discussed which of the
remaining scan points yield a cosmological relic density in the allowed
range $\Omega_\chi h^2 \le 0.3$, and which of these are in the preferred
range $\Omega_\chi h^2 \ge 0.1$. We exclude from further consideration
the over-dense points with $\Omega_\chi h^2 > 0.3$, and rescale the
predicted cross sections for
under-dense points with $\Omega_\chi h^2 < 0.1$ as in
(\ref{halofraction}).

\begin{figure}[htb]
\vskip -3cm
\hspace*{-.60in}
\begin{minipage}{8in}
\epsfig{figure=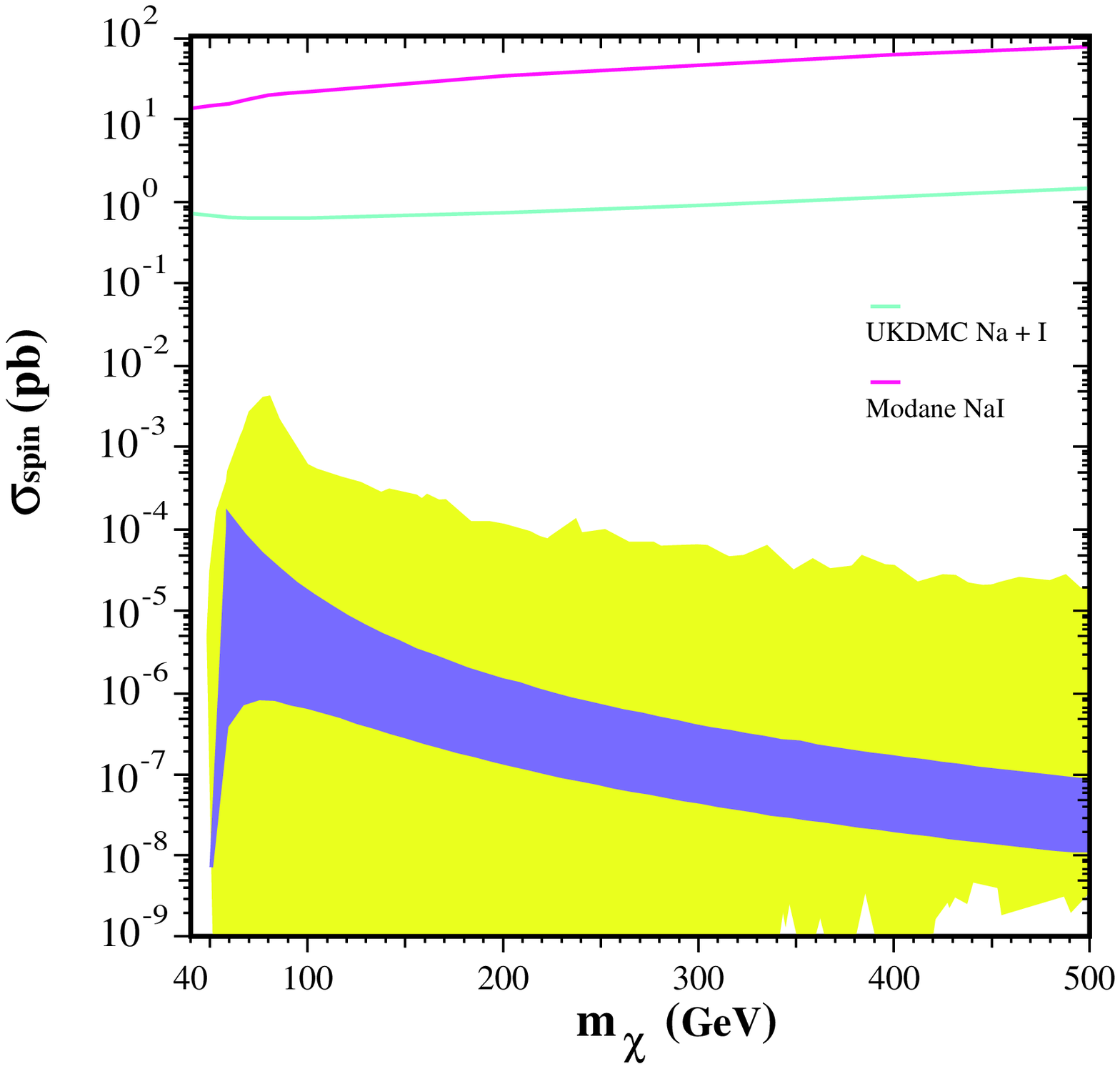,height=5.5in}
\hspace*{-.70in}\epsfig{file=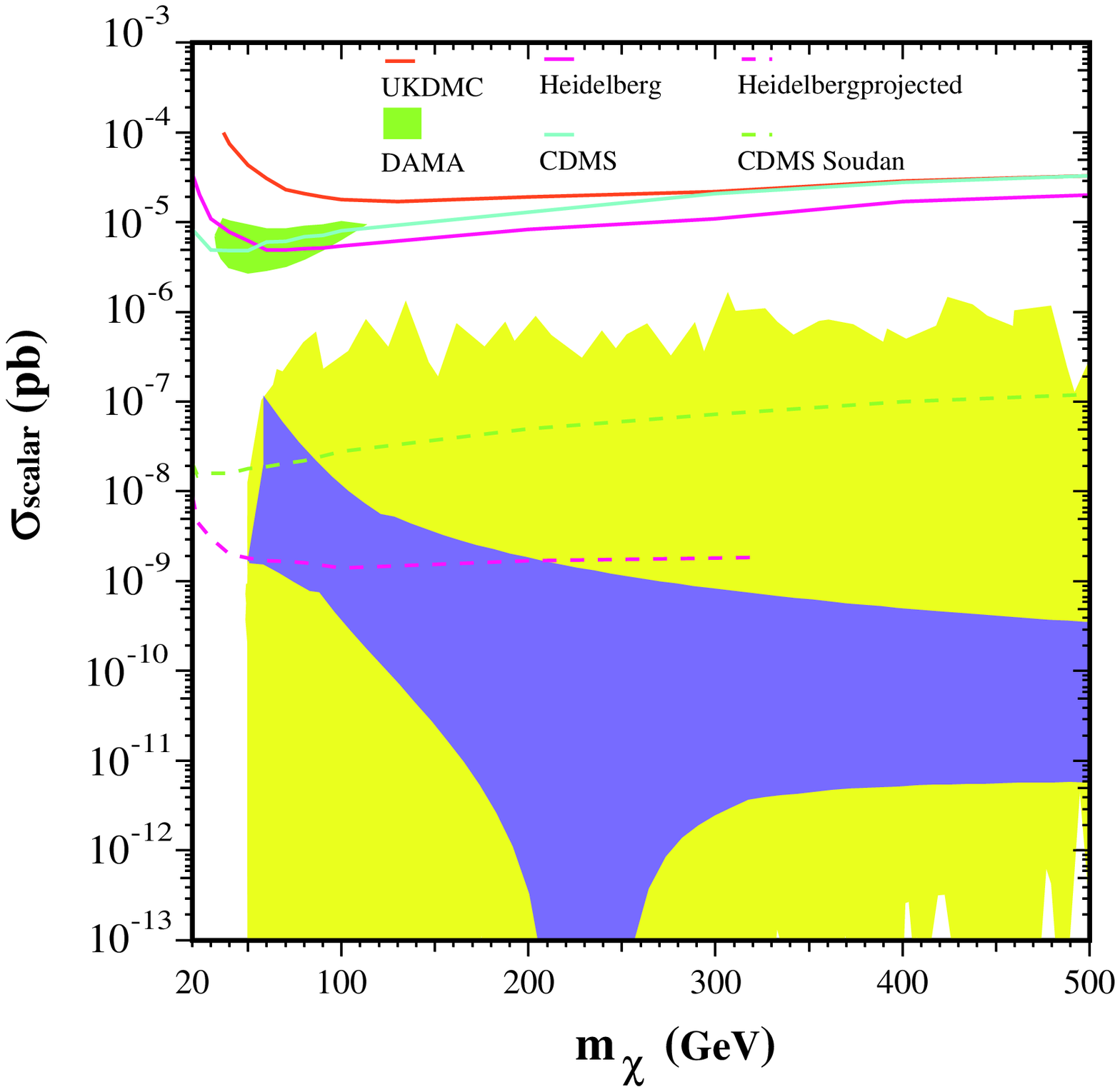,height=5.5in}
\hfill
\end{minipage}
\caption{\it
Compilations of our allowed ranges
for (a) the spin-dependent elastic neutralino-proton
cross section, and (b) the spin-independent elastic neutralino-proton
cross section for both the values of
$\tan\beta$ and the signs of $\mu$ studied.
Our results are compared in panel (a) with the available
experimental upper limits~\cite{Gait}, and in panel (b) with the detection
reported by the DAMA Collaboration~\cite{DAMA}, as well as with
upper limits from other experiments~\cite{Gait}.}
\label{fig:summary}
\end{figure}

The cross sections we predict for spin-dependent and spin-independent
elastic neutralino-proton scattering for different values of $\tan\beta$ 
and the sign of $\mu$ are shown in Figs.~\ref{fig:spin} and
\ref{fig:scalar}, respectively. We provide in Fig.~\ref{fig:summary}
a compilation of our results, compared with the present experimental
upper limits on the cross sections~\cite{Gait} and the detection of
spin-independent scattering reported by the DAMA
Collaboration~\cite{DAMA}. 
We see that our predicted cross sections are
well below the experimental upper limits for both
the spin-dependent and -independent cases. We are unable to find MSSM
parameter sets consistent with our relaxed universality assumptions that
come close to explaining the DAMA measurements. Our assumptions
would need to be questioned if the neutralino scattering interpretation
of the DAMA data is confirmed. For example, we have restricted our
attention to models with $\tan \beta \le 10$.
Alternatively, the DAMA data might favour models with values of
$m_{\tilde q} / m_{\tilde \ell}$, obtained either by relaxing the
input universality assumption, or by imposing it at some renormalization
scale below the conventional supersymmetric GUT scale~\cite{lowuni}.

In the future, we plan to improve the available relic density calculations
by extending them to larger $\tan \beta$ and incorporating consistently
all coannihilation processes. On the experimental side, we expect that
other Collaborations will soon be able to confirm or exclude definitively
the DAMA interpretation of their annual modulation signal as being due to
neutralino scattering. Looking further ahead, we interpret our results as
indicating a high priority for a new generation of direct dark matter
detection experiments~\cite{Genius} with a much higher sensitivity.

\
\vskip 0.5in
\vbox{
\noindent{ {\bf Acknowledgments} } \\
\noindent 
We thank Toby Falk and Gerardo Ganis for many related discussions. 
The work of K.A.O. was supported in part by DOE grant
DE--FG02--94ER--40823.}


\newpage

\begin{thebibliography}{99}

\bibitem{triangle}
N.~Bahcall, J.~P.~Ostriker, S.~Perlmutter and P.~J.~Steinhardt,
Science {\bf 284} (1999) 1481.

\bibitem{GJK}
For a review, see, e.g., G. Jungman, M. Kamionkowski and K. Griest, Phys.
Rep.
{\bf 267} (1996) 195.

\bibitem{GW}
M.W. Goodman and E. Witten, Phys. Rev. {\bf D31} (1986) 3059.

\bibitem{EHNOS}
J. Ellis, J.S. Hagelin, D.V. Nanopoulos, K.A. Olive
and M. Srednicki, Nucl. Phys. {\bf B238} (1984) 453.

\bibitem{MSSM}
For reviews, see:
H.P. Nilles, Phys. Rep. {\bf 110} (1984) 1;
H.E. Haber and G.L. Kane, Phys. Rep. {\bf 117} (1995) 75.

\bibitem{EFlO1}
J.~Ellis, A.~Ferstl and K.~A.~Olive,
Phys. Let.. {\bf 481} (2000) 304.

\bibitem{EF}
J. Ellis and R. Flores, Nucl. Phys. {\bf B307} (1988) 883;
Phys. Lett. {\bf B263} (1991) 259; Phys. Lett. {\bf B300} (1993) 175.

\bibitem{etal}
K. Griest, Phys. Rev. {\bf D38} (1988) 2357;
R. Barbieri, M. Frigeni and G. Giudice, Nucl. Phys. {\bf B313} (1989)
725; R. Flores, K.A. Olive and M. Srednicki, Phys. Lett. {\bf B237}
(1990) 72; M. Drees and M. M. Nojiri, Phys. Rev. {\bf D48} (1993) 3483; V.
Bednyakov, H.V. Klapdor-Kleingrothaus and S. Kovalenko, Phys. Rev. {\bf
D50} (1994) 7128; H. Baer and M. Brhlik Phys. Rev. {\bf D57}
(1998) 567.

\bibitem{leut}
H. Leutwyler, hep-ph/9609465.

\bibitem{Cheng}
H.-Y. Cheng, Phys. Lett. {\bf B219} (1989) 347.

\bibitem{Gasser}
J. Gasser, H. Leutwyler, and M. E. Sainio, Phys. Lett. {\bf B253} (1991)
252; M.~Knecht, hep-ph/9912443.

\bibitem{Mallot}
G. Mallot, hep-ex/9912040.

\bibitem{Gait}
R. Gaitskell and V. Mandic, {\tt http://cdms.berkeley.edu/limitplots/}.

\bibitem{bigguys}
B. Sadoulet,
Nucl.\ Phys.\ Proc.\ Suppl.\  {\bf 77} (1999) 389;
A. Bottino, F. Donato, N. Forengo and S. Scopel,
Phys.\ Rev.\  {\bf D59} (1999) 095003;
Phys.\ Rev.\  {\bf D59} (1999) 095004;
Astropart.\ Phys.\  {\bf 10} (1999) 203;
Astropart.\ Phys.\  {\bf 13} (2000) 215;
hep-ph/0001309.

\bibitem{lowguys}
A.~Corsetti and P.~Nath,
hep-ph/0003186.

\bibitem{DAMA}
R.~Bernabei {\it et al.}, DAMA Collaboration,
Phys.\ Lett.\  {\bf B480} (2000) 23.

\bibitem{CDMS}
R.~Abusaidi {\it et al.}, CDMS Collaboration,
astro-ph/0002471.

\bibitem{EFGOS}
J. Ellis, T. Falk, K.A. Olive and M. Schmitt,
Phys.\ Lett.\ {\bf B388} (1996) 97
and Phys.\ Lett.\ {\bf B413} (1997) 355;
J.~Ellis, T.~Falk, G.~Ganis, K.A.~Olive
and M.~Schmitt,
Phys.\ Rev.\ {\bf D58} (1998) 095002.

\bibitem{EFGO}
J.~Ellis, T.~Falk, G.~Ganis and K.~A.~Olive, Phys.\ Rev.\ {\bf D} (in
press),  hep-ph/0004169.

\bibitem{arno}
E.~Accomando, R.~Arnowitt, B.~Dutta and Y.~Santoso,
hep-ph/0001019;
R.~Arnowitt, B.~Dutta, and Y.~Santoso, hep-ph/0005154.


\bibitem{lowuni} E. Gabrielli, S. Khalil, C. Munoz, E. Torrente-Lujan,
hep-ph/0006266.

\bibitem{FFO1}
T. Falk, A. Ferstl and K.A. Olive, Phys. Rev. {\bf D59} (1999) 055009.

\bibitem{FFO2}
T. Falk, A. Ferstl and K.A. Olive, Astropart. Phys. {\bf 13} (2000) 301,
hep-ph/9908311.

\bibitem{CIN}
U.\ Chattopadhyay, T.\ Ibrahim and P.\ Nath,  Phys. Rev. {\bf D60}
(1999) 063505.

\bibitem{fg} P. Gondolo and K. Freese, hep-ph/9908390; S.Y. Choi, 
hep-ph/9908397

\bibitem{BOOM}
P.~de Bernardis {\it et al.},
Nature {\bf 404} (2000) 955;
A.~E.~Lange {\it et al.},
astro-ph/0005004.

\bibitem{MAXI}
A.~Balbi {\it et al.},
astro-ph/0005124.

\bibitem{ES}
J.~Ellis and P.~Sikivie,
Phys.\ Lett.\  {\bf B321} (1994) 390.

\bibitem{McOS}J. McDonald, K. A. Olive and M. Srednicki,  Phys. Lett.
 {\bf B283} (1992) 80.

\bibitem{gs}K. Griest and D. Seckel, Phys. Rep. {\bf D43} (1991) 3191.

\bibitem{co2} S. Mizuta and M. Yamaguchi, Phys. Lett. {\bf B298} (1993)
120.

\bibitem{EFOSi}
J.~Ellis, T.~Falk and K.A.~Olive,
Phys.\ Lett.\ {\bf B444} (1998) 367;
J.~Ellis, T.~Falk, K.A.~Olive and M.~Srednicki, Astropart. Phys. {\bf 13}
(2000) 181.

\bibitem{BDD}
C.~Boehm, A.~Djouadi and M.~Drees,
hep-ph/9911496.

\bibitem{FMW}
J.~L.~Feng, K.~T.~Matchev and F.~Wilczek,
Phys.\ Lett.\  {\bf B482} (2000) 388.

\bibitem{newLEP}
The latest LEP limits on sparticles and Higgs bosons may be found in links
on \\
{\tt
http://delphiwww.cern.ch/$\tilde{}$offline/physics{\_}links/lepc.html}.

\bibitem{newHiggs}
M.~Carena, S.~Heinemeyer, C.E.~Wagner and G.~Weiglein,
hep-ph/9912223 and references therein.

\bibitem{bsg}
P. Nath and R. Arnowitt, Phys.\ Rev.\ Lett.\ {\bf 74} (1995) 4592;
F.~M.~Borzumati, M.~Drees and M.~M.~Nojiri,
Phys.\ Rev.\  {\bf D51} (1995) 341;
H. Baer and M. Brhlik, Phys.\ Rev.\ {\bf D55} (1997) 3201.

\bibitem{AF}
H. Baer, M. Brhlik and D. Casta\~no, Phys.\ Rev.\ {\bf D54} (1996) 6944;
S. Abel and T. Falk, Phys.\ Lett.\ {\bf B444} (1998) 427.

\bibitem{up} K.A. Olive and M. Srednicki, Phys. Lett. {\bf B230}
 (1989) 78;
 Nucl. Phys. {\bf  B355} (1991) 208; K. Greist, 
M. Kamionkowski and M.S. Turner, Phys. Rev.
{\bf D41} (1990) 3565.

\bibitem{Genius}
L.~Baudis {\it et al.}, GENIUS Collaboration,
hep-ph/9910205.


\end{thebibliography}
\end{document}